\documentclass[a4paper,11pt]{article}
\pdfoutput=1 

\usepackage{jheppub} 

\usepackage[T1]{fontenc} 
\usepackage{bbm}
\usepackage{mycommands}

\title{\boldmath QGP modification to single inclusive jets in a calibrated transport model}

\author[a,b,1]{Weiyao Ke}
\author[b,c,2]{Xin-Nian Wang}

\affiliation[a]{Department of Physics, University of California, Berkeley\\
366 LeConte Hall, Berkeley CA 94720, United States}

\affiliation[b]{Nuclear Science Division, MS 70R0309\\
Lawrence-Berkeley National Laboratory\\
1 Cyclotron Rd, Berkeley CA 94720, United States}

\affiliation[c]{Key Laboratory of Quark and Lepton Physics (MOE) and Institute of Particle Physics, \\
Central China Normal University, Wuhan 430079, China}

\emailAdd{weiyaoke@lbl.gov}
\emailAdd{xnwang@lbl.gov}
\abstract{
We study inclusive jet suppression and modifications in the quark-gluon plasma (QGP) with a transport-based model.
The model includes vacuum-like parton shower evolution at high-virtuality, a linearized transport for jet-medium interactions, and a simple ansatz for the jet-induced hydrodynamic response of the medium.
Model parameters are calibrated to nuclear modification factors for inclusive hadron $R_{AA}^{h}$ and single inclusive jets $R_{AA}^{j}$ with cone size $R=0.4$ in 0-10\% central Au-Au and Pb-Pb collisions measured at the RHIC and LHC.
The calibrated model consistently describes the cone-size dependent $R_{AA}^{j}(R)$, modifications to inclusive jet fragmentation functions and jet shape.
We discuss the origin of these modifications by analyzing the medium-induced jet energy flow in this model and elucidate the interplay of hard parton evolution and jet-induced medium response.
In particular, we demonstrate that the excess of soft hadrons at $p_T\sim 2$ \GeVc in jet fragmentation function and jet shape at large $r=\sqrt{\Delta \eta^2+\Delta \phi^2}$ are consequences of both soft medium-induced gluon radiation and jet-induced medium excitation.
}

\begin{document} 
\maketitle
\flushbottom

\section{Introduction}
\label{sec:intro}
Jets produced in nuclear collisions probe the properties of the color-deconfined quark-gluon plasma (QGP).
Strong final-state jet-medium interactions suppress the yield of jets as well as high-$p_T$ hadrons.
This phenomenon, known as the ``jet quenching'', is among the key signatures of the formation of the QGP~\cite{PhysRevD.33.717,GYULASSY1990432,PhysRevLett.68.1480,Qin:2015srf} and is also used to determine the stopping power of QGP to colored particles~\cite{Bass:2008rv,PhysRevC.90.014909,Andres:2016iys,Xu:2017obm,Xie:2019oxg}.
Besides yield suppression, measurements of modifications to jet constituents and internal structures become available in recent years at both the RHIC and the LHC energies~\cite{Connors:2017ptx}.
It is therefore important to develop a consistent theoretical picture to understand the suppression patterns of jets and hadrons and the modified jet internal structure \cite{Chien:2015hda,Casalderrey-Solana:2015vaa,Kang:2016ehg,Chang:2016gjp,Tachibana:2018yae,Casalderrey-Solana:2018wrw,Pablos:2019ngg,Caucal:2020xad,Chen:2020tbl}, which requires the modeling of microscopic parton-medium interactions.

Conceptually, interactions between energetic jet partons and medium constituents fall into two categories:
elastic collisions between jet and medium partons and the medium-induced parton splittings.
The formula for a single medium-induced parton splitting \cite{Baier:1994bd,BAIER1997265,Zakharov:1996fv,Zakharov:1997uu,Baier:1998kq,Wiedemann:2000za,Gyulassy:2000er,Arnold:2002ja} have been known for decades, and energy loss due to elastic collisions has also been studied in both weakly-coupled \cite{THOMA1991128,PhysRevD.44.R2625,Mustafa:2003vh,Mustafa:2004dr,Djordjevic:2006tw,Wang:2006qr,Schenke:2008gg,PhysRevLett.100.072301} and strongly-coupled field theory \cite{Liu:2006ug}.
Applying these theoretical inputs to jet phenomenology in the dense QGP medium, one needs to include multiple interactions, which has led to the use of different evolution equations.
For example, many studies use DGLAP-like QCD evolution equations with medium-modified evolution kernels or equations derived from high energy effective theory including medium effects ~\cite{Wang:2002ri,Idilbi:2008vm,Ovanesyan:2011xy,Chien:2015vja,Cao:2017qpx}.
In this work, we take another widely used approach based on transport equations from the kinetic theory \cite{Arnold:2002zm,Arnold:2003zc,Jeon:2003gi,Schenke:2009gb,Ghiglieri:2015ala,He:2015pra,Cao:2017hhk,He:2018xjv}.
In this approach, multiple radiations and collisions are included as the time-evolution of a rate equation as governed by the Boltzmann-type transport equation.
The first goal of this work is to develop a simulation procedure around a previously developed linearized partonic transport model {\tt LIDO} \cite{Ke:2018jem}.
This simulation procedure will be referred to as ``a transport-based model'' in this paper.

Given the premise of using a transport equation for the jet-medium interaction, we are confronted with two problems in the modeling.

The first problem concerns with the scale evolution of the parton showers from the hard scale $Q\sim \pThard$ to $Q_0 \gtrsim \LambdaQCD$.
Because jets have scale-dependent partonic contents, one has to choose a matching scale where the transport equation in medium starts to apply.
Many studies, such as those used the Linearized Boltzmann Transport (LBT) model \cite{He:2015pra,Cao:2017hhk,Chen:2017zte,He:2018xjv}, included the full vacuum shower before applying the transport method, i.e., the transition was made at the same scale as the minimum evolution scale in p-p collisions.
The study using the MARTINI model explored two scenarios \cite{Schenke:2009gb}. 
One includes the full evolution of the vacuum parton shower before the onset of medium effects and the other uses a formation-time argument to motivate the choice of matching scale $Q_0 \sim \sqrt{p_T/\tau_0}$, where $\tau_0$ is the initial proper time of hydrodynamic evolution of the medium.
It was found that different matching procedures affect the phenomenological choice of the effective jet-medium coupling constant.
In a series of recent works~\cite{PhysRevLett.120.232001,Caucal:2020xad}, the authors carefully explored the phase-space boundaries between vacuum-like and medium-induced emissions in a finite and static medium.
The matching scale between high-virtually vacuum-like emissions and the medium-induced emissions are motivated by the momentum broadening $Q_0\sim (\hat{q}E)^{1/4}$ obtained using the multiple-soft collision approximation.
The aforementioned studies assume the scale evolution of the parton is either unmodified by the medium or only restricted in emission phase-space.
Other models, such as the MATTER event generator~\cite{Majumder:2013re}, include medium-modified splitting functions into the DGLAP-type evolution equation according to a higher twist approach \cite{Wang:2001ifa,Majumder:2009ge,Wang:2009qb}.
The MATTER model has been combined with either LBT or MARTINI transport models in the JETSCAPE framework \cite{Cao:2017zih,Kauder:2018cdt,Putschke:2019yrg}.
The dependence of model predictions on choices of $Q_0$ was studied \cite{Cao:2017zih,Soltz:2019aea,Park:2020mkl}.

In this work, we use the simpler method that the parton  evolution above a certain scale $Q_0$ is unmodified by the medium and the evolution below $Q_0$ is modeled a time evolution governed by the transport equation.
Despite $Q_0$ is treated as a free parameter in our model, we expect the physical value of $Q_0$ should be comparable to the momentum broadening $\sqrt{\langle k_\perp^2\rangle}$ acquired during the formation time of an emission in the QGP medium\footnote{
For multiple-soft collision approximation of jet-medium interaction in a static medium, this is $Q_0\sim (\hat{q}E)^{1/4}$.}.
We will argue that in a fast expanding medium, typical values of $\langle k_\perp^2\rangle$ are relatively insensitive to the parton energy and are mainly determined by medium properties.
Therefore, it is reasonable to use a single matching scale $Q_0$ for all partons.

For partons going out of the QGP fireball, QGP modifications vanishes and we apply the vacuum-like evolution again, neglecting jet interactions with the hadronic matter.
However, one should note that this procedure is contradictory to the current cross-over picture of the phase transition at small baryon chemical potential, where medium properties are continuous across $T_c$.
Moreover, a recent study indicates substantial interactions between high-$p_T$ pions and the hadronic matter near $T_c$ \cite{Dorau:2019ozd}.
We will leave the study of jet modifications in the hadronic phase to future works.

The second problem is that a transport equation is only defined for well-defined quasiparticles.
In the QGP medium, only hard particles with energy much greater than the temperature of the medium ($E\gg T$) can be identified as good quasiparticles, including both hard partons from partons showers as well as recoiled medium partons that acquires a large momentum in the collision with a jet parton.
However, the change in the dynamics of the soft medium in response to the passing of a jet is beyond the scope of the linearized partonic transport equation with weakly-coupled interactions.
The perturbed medium consists of particles with thermal momentum $p\sim \mathcal{O}(T)$ and its dynamics are better modeled in a hydrodynamic approach.
Models that couple a transport model to hydrodynamic \cite{Chen:2017zte,Tachibana:2017syd}, models based on instant thermalization \cite{Casalderrey-Solana:2016jvj} and linearized hydrodynamic response \cite{Casalderrey-Solana:2020rsj}, and methods of including medium recoil particles \cite{He:2015pra,KunnawalkamElayavalli:2017hxo} have been used to address such a problem.
In this work, we will develop a simple ansatz to model the hydrodynamic-like response of medium, including the freeze-out of particles.
We use a single term ``jet-induced medium response'' to refer to both the hydrodynamic-like response of soft medium dynamics and the hard recoiled medium partons.

We will introduce this transport-based model in detail in section \ref{sec:model}, including a review of the {\tt LIDO} transport model and our approach for addressing the above-mentioned problems.

For reliable predictions with quantified uncertainty, we will perform a Bayesian model calibration of the model parameters to the inclusive jet and hadron suppressions in central nuclear collisions at both the RHIC and the LHC.
This also tests the model's ability to describe both jet and hadron suppression simultaneously.

In section \ref{sec:energy-flow}, using the calibrated model, we analyze the energy flow induced by jet-medium interactions in the two-dimensional space $(\zjet, r)$, where $\zjet$ is the longitudinal momentum fraction of particles relative to the momentum of the jet,
and $r$ is the radial distances between particles and jets
\begin{eqnarray}
\zjet &=& \frac{p\cdot \pjet}{(\pTjet)^2} \approx \frac{p_T\cos(r)}{\pTjet}, \\
r &=& \sqrt{\left(\phi-\phijet\right)^2+\left(\eta-\etajet\right)^2}.
\end{eqnarray}
Mapping out this energy flow and differentiating effects from different transport mechanisms help to interpret our prediction on jet modifications.
Qualitatively, medium-induced radiations predominately transfer energy from high-$\zjet$ to low-$p_T$ particles, while elastic collisions and jet-induced medium response carry energy to large angles.

In section \ref{sec:prediction}, we predict modifications to jet fragmentation functions and jet shape as well as the cone-size dependence of jet suppression using the calibrated model.
We try to identify different energy transport mechanisms in these modifications, which provides a deeper understanding of jet-medium interaction.

Finally, we summarize in section \ref{sec:conclusion} our results and comment on the jet transport parameter determined in this work.

\section{A transport-based model for jet evolution in nuclear collisions}
\label{sec:model}
In this section, we introduce the jet evolution model that has been outlined in the introduction.
It involves three stages of evolution:
\begin{enumerate}
\item The initial hard processes of jet production and vacuum-like evolution of parton shower from $Q\sim \pThard$ to $Q=Q_0$ is modeled by {\tt Pythia8}.
\item The initial jet shower parton production is followed by partons-medium interactions, which are treated in a previously developed linearized partonic transport model {\tt LIDO}.
Meanwhile, jet partons excite the medium in two ways: medium partons that acquire a hard recoiled momentum in collisions with jet partons; the perturbed soft medium partons seed a linearized hydrodynamic-like response.
The subsequent evolution of the hard recoil partons is also treated by the transport equation.
Together, hard recoils and the hydrodynamic response are termed as the jet-induced medium excitation.
\item A stage of vacuum evolution of parton shower from $Q_0$ to $\LambdaQCD$ occurs outside the QGP medium.
It is followed by hadronization and particle decays.
This stage is again modeled by {\tt Pythia8}.
\end{enumerate}
We will discuss models for each stage in detail and the matching procedures between them.

In addition, the soft QGP medium evolution is described by a 2+1D viscous hydrodynamics with event-averaged initial conditions. 
Due to the lack of event-by-event fluctuations and longitudinal evolution, we will only focus on midrapidity observables in most central nuclear collisions at the RHIC and the LHC in this paper.
Details of the medium evolution model can be found in appendix \ref{app:medium}.

\subsection{Jet evolution in vacuum and initialization of transport equation}
\label{sec:model:vac}
We use the {\tt Pythia8} event generator~\cite{Sjostrand:2014zea} with ATLAS A14 central tune~\cite{ATL-PHYS-PUB-2014-021} to sample initial hard scatterings and generate the initial parton showers.
The proton partron distribution function (PDF) in {\tt Pythia8} takes the leading order CTEQL1 PDF \cite{Pumplin:2002vw}, and the nuclear PDF takes the leading-order EPS09 parametrization~\cite{Eskola:2009uj}.

The initial parton radiations are ordered in decreasing virtuality, which is just the transverse momentum of the radiated particle in this case.
In proton-proton collisions, the scale $Q$ starts from that of the initial hard collision $\pThard$ down to a cut-off scale, which is close to $\LambdaQCD$.
In nuclear collisions, we have argued that the high-virtuality vacuum-like evolution should be stopped scale $Q_0$, comparable to the virtuality of partons acquired in the medium.
The latter can be estimated in a parton splitting process including medium effects of momentum broadening: a parton with energy $E$ radiates an almost collinear parton with energy $xE$ and transverse momentum $k_\perp\sim Q$ relative to the original parton.
Within the radiation formation time
\begin{eqnarray}
\tau_f = \frac{2x(1-x)E}{k_\perp^2},
\end{eqnarray}
collisions with medium broaden the transverse momentum to,
\begin{eqnarray}
\langle \Delta k_\perp^2 \rangle = \int_{t_0}^{t_0+\tau_f} \qhat(E) dt,
\end{eqnarray}
where $\qhat=d\langle \Delta k_\perp^2 \rangle/dt$ is the jet transport parameter.
Medium impacts are small when $Q^2\gg \Delta k_\perp^2$, where we apply the vacuum-like evolution.
For smaller $Q^2$, the typical transverse momentum saturates at $\sqrt{\Delta k_\perp^2}$ and the vacuum-like parton shower gives way to the in-medium evolution governed by the transport equation.

In the introduction, we have summarized different choices of $Q_0$ in the past studies, including the use of fixed $Q_0$ and parton-energy-dependent choices, such as $Q_0\propto E^{1/2}$ obtained using formation time argument and $Q_0\propto E^{1/4}$ motivated by the momentum broadening in a static medium due to multiple-soft collisions.
Here, we argue that because of the fast dropping of medium temperature due to expansion, $Q_0\sim \sqrt{\Delta k_\perp^2}$ is relatively independent of the parton energy.
In section \ref{sec:bayes:posterior}, We will verify this argument by running the model in a medium simulated by the hydrodynamics and show dependence of the distribution of $\Delta k_\perp^2$ on the parton energy. 
Right now, we first demonstrate this argument using a medium under Bjorken expansion \cite{PhysRevD.27.140}.
In a medium with Bjorken flow solution $\hat{q}\propto T^3 \propto \tau^{-1}$ or $\qhat(\tau)\tau \approx \qhat(\tau_0)\tau_0$.
The momentum broadening is estimated from the self-consistent relation
\begin{eqnarray}
\langle \tau_f^{-1}\rangle \equiv \frac{\langle k_\perp^2 \rangle}{2x(1-x)E} \approx \frac{\int^{\tau_f} \qhat d\tau}{2x(1-x)E} = \frac{\qhat_0\tau_0 \ln\frac{\tau_f}{\tau_0}}{2x(1-x)E}
\end{eqnarray}
Approximately, $\langle k_\perp^2\rangle \approx \qhat(\tau_0)\tau_0 \ln (E/\qhat(\tau_0)\tau_0^2)$ and it only increases logarithmically with the parton energy. 
Therefore, at given beam energy, collision system, and centrality, it is reasonable and also practical to use a single $Q_0$ scale to transit the entire parton shower from vacuum-like scale evolution to parton transport in the medium.

Tuning $Q_0$ parameter is effectively changing the minimum virtuality scale in {\tt Pythia8} when we generate the initial parton shower.
Specific to this study, we treat $Q_0$ values in central Pb-Pb collisions at $\sqrtsNN=5.02$ TeV and in central Au-Au collisions at $\sqrtsNN=200$ GeV as independent parameters: $\QLHC$ and $\QRHIC$.
In section \ref{sec:bayes}, after their values are calibrated to experimental data, we perform a consistency check by comparing the distribution of momentum broadening in our simulations to the extracted values of $\QLHC$ and $\QRHIC$.

Finally, to initialize the transport equation, which requires both the momentum and spatial information, we assign space-time coordinates for partons created in the initial parton shower.
First, a transverse position $\bfxperp$ is sampled according to the density of binary collisions in the transverse overlap region. 
The coordinates $x^\mu=(0, \bfxperp, 0)$ are assigned to the hardest partons directly created from the hard collision.
Consider parton $a$ with energy $E$ and coordinate $x_a$ splits into partons $b$ and $c$, and parton $b$ has a larger energy $xE$ than parton $c$, $x>1/2$.
We assume that the radiation of the softer parton is delayed by its formation time $\tau_f$ and is unmodified by the medium.
The production space-time coordinate of parton $c$ is
\begin{eqnarray}
x_c^\mu &=& x_a^\mu +\frac{p_c^\mu}{p_c^0} \tau_f.
\end{eqnarray}

\subsection{The {\tt LIDO} transport model}
\label{sec:model:lido}
The jet shower partons at scale $Q_0$ provide the initial condition for the transport equation.
We use a previously developed linearized partonic transport model {\tt LIDO} \cite{Ke:2018jem} for this stage.
This model relies on several assumptions:
\begin{itemize}
\item The first assumption is the weakly-coupled interactions in which the coupling constant $\alpha_s$ or $g_s=\sqrt{4\pi \alpha_s}$ are parametrically small.
In this limit, jet partons interact with medium quasiparticles via screened gluon exchanges.
We use the three-flavor leading-order running coupling constant with $\LambdaQCD=0.2$ GeV. 
At finite temperature, we assume that the running scale has to be greater than a medium scale $\mumin\propto T$.
This results in the following ansatz of running coupling at finite temperature,
\begin{eqnarray}
\alpha_s(Q, T) = \frac{4\pi}{9} \frac{1}{\ln \left(\frac{\max\{Q^2,\mumin^2\}}{\LambdaQCD^2}\right)}.
\end{eqnarray}

\item The linearization assumption: the equation only treats hard particles with $p\gg T$, where the population is small compared to thermal partons in the medium. 
We also neglect collisions between hard partons and their interactions with the jet-induced medium excitation.
Therefore, the transport equation is linearized with respect to hard partons.
\item The final assumption requires the background medium is composed of massless partons in local thermal equilibrium. Thermal partons are then sampled from Boltzmann-type distribution function $f_0(p)=e^{-p\cdot u/T}$, where $u,T$ are the flow four-velocity and temperature of the local medium obtained, respectively, from the hydrodynamics simulation.
\end{itemize}
Given these premises, a Boltzmann-type transport equation evolves the hard-parton distribution function $f(t, \bfx; \bfp)$
\begin{eqnarray}
p\cdot\partial [f \Theta(p\cdot u-\Emin)] = (p\cdot u) \Theta(p\cdot u-\Emin) \left\{ \mathcal{D} +\mathcal{C}_{1\leftrightarrow 2} +\mathcal{C}_{2\leftrightarrow 2} +\mathcal{C}_{2\leftrightarrow 3}\right\}.
\label{eq:lido}
\end{eqnarray}
The left-hand side of the above equation is the ballistic transport term. 
We have inserted step functions to emphasize that the transport equation only handles hard partons with energy greater than $\Emin$ in the co-moving frame.
Throughout this work, $\Emin = 4T$, which is slightly larger than the average kinetic energy of a thermal parton.

\paragraph{Elastic jet-medium interactions}
Elastic interactions refer to the processes that conserve the number of hard partons.
In the {\tt LIDO} model, elastic collisions that only involve small-momentum transfer $q_\perp < \Qcut$ transverse to the hard parton are included in drag and diffusion operator $\mathcal{D}$,
\begin{eqnarray}
\mathcal{D}[f] = \left(\eta_D p + \frac{\qhat_S}{4}\frac{\partial^2}{\partial p^2}\right)f.
\end{eqnarray}
Here, the $\qhat_S$ denote the transport parameter due to soft interactions \cite{Ghiglieri:2015ala}
\begin{eqnarray}
\label{eq:qhat_soft}
\qhat_{a,s}(p, T) =  6\pi T^3 C_R \int_{0}^{\Qcut^2} \frac{\alpha_s^2(q_\perp)}{q_\perp^2+m_D^2} dq_\perp^2,
\end{eqnarray}
where $a$ represents quarks or gluons. 
The respective Casimir factors are $C_A=3$ for gluons ($a=``g"$) and $C_F=4/3$ for quarks ($a=``q"$).
$m_D=\sqrt{3/2}g(\mumin)T$ is the Debye screening mass of the QGP with three flavors of light quarks and
$\eta_D$ is the momentum drag coefficient determined by the  fluctuation-dissipation relation $\eta_D = \qhat_S/(4ET)$.
We set $\Qcut=\min\{2 m_D, \sqrt{6ET}\}$ throughout this work. 
Here $\sqrt{6ET}$ estimates the average center-of-mass energy between jet partons and medium partons\footnote{Separating small-$q$ collisions into a diffusion equation seem to be an unnecessary step if we only want to include leading order weakly-coupled physics. But a diffusion formulation has a different range of applicability than matrix-element based collisions. For example, one can easily implement diffusion coefficients from strongly-coupled theories for the soft-momentum exchange process.}.

Elastic collisions that involve large momenta transfer $q_\perp > \Qcut$ are included explicitly in a two-body collision term $C_{2\leftrightarrow 2}$ for particle specie $a$,
\begin{eqnarray}
\nonumber
C_{2\leftrightarrow 2}^a(p_a) &=& \sum_{bcde} \nu_c\int_{q_\perp>\Qcut} d\Pi^{p_b p_c}_{p_d p_e} |M|^2_{bc;de}f^b(p_b)f_{0}(p_c)\\
&&
(2\pi)^3\left[
-\delta^{(3)}(p-p_b)\delta^{ab}
+\sum_{i=d,e}\delta^{(3)}(p-p_i)\delta^{ai}
\right]
\label{eq:C22}
\end{eqnarray}
where $p_b, p_c$ and $p_d, p_e$ are four momenta of the initial-state partons and final-state partons of the collision, respectively. 
$q=p_b-p_d$ is the four-momentum change between the hardest initial-state and final-state particle in the co-moving frame.
$|M(q)|^2_{bc,de}$ is the squared amplitude of the two-body collision in the vacuum\footnote{The difference between screened propagators and propagators in the vacuum is assumed to be small when $\Qcut\gg m_D$ \cite{Ghiglieri:2015ala}, rationalizing the use of vacuum matrix-elements.}, averaging over initial-state and summing over final-state spins and degeneracies.
The running coupling constant is evaluated at scale $\sqrt{-q^2}$.
$f^b(p_b)$ is the distribution function of the initial-state hard parton, and $f_0(p_c)$ is the equilibrium distribution function of the initial-state medium parton.
We first integrate over the entire phase space of the initial-state $\{p_i\}$ and final-state $\{p_f\}$ particles $\int_{q_\perp>\Qcut}d\Pi_{\{p_i\}; \{p_f\}}$, restricted to hard-momentum transfers $q_\perp>\Qcut$
\begin{eqnarray}
d\Pi^{\{p_i\}}_{\{p_f\}} = (2\pi)^4\delta^{(4)}\left(\sum_{l\in\{p_i\}}l^\mu-\sum_{l\in\{p_f\}}l^\mu\right)
\prod_{l\in \{p_i\} \{p_f\}} \frac{dl^3}{2l^0 (2\pi)^3}.
\end{eqnarray}
Then, we insert in the second line of equation \ref{eq:C22} a set of $\delta$-functions in momentum and particle specie projectors to project out contributions to the distribution function of hard parton species $a$.
The first negative term in the projector corresponds to the loss term, and the second summation that goes over the final-state particles corresponds to the gain term.
Finally, the expression sums all reaction channels, and $\nu_c$ is the degeneracy factor of the medium parton $c$, including spins, colors, and the three light flavors: up, down, and strange.

Because of the separation in $q_\perp$, the jet transport parameter for parton specie $a$ is also composed of a soft part and a hard part,
\begin{eqnarray}
\qhat_{a}(p, T) = \qhat_{a,s} + \qhat_{a,h},
\end{eqnarray}
where $\qhat_{a,s}$ has been defined in Equation \ref{eq:qhat_soft}, and we get $\qhat_{a,h}$ by computing the average momentum broadening due to large-$q$ scatterings per unit time
\begin{eqnarray}
\qhat_{a,h}(p_a) &=& \sum_{b,c,d,e}\nu_c\int_{q_\perp>\Qcut} d\Pi^{p_bp_c}_{p_dp_e}  q_\perp^2 |M|^2_{p_b p_c; p_d p_e}f_0(p_c) (2\pi)^3\delta^{(3)}(p-p_a)\delta^{ab}\delta^{ad}.
\label{eq:qhat_hard}
\end{eqnarray}

\paragraph{Medium-induced parton radiation} 
Medium-induced parton radiations change the number of hard partons in collisions.
Similar to the separation for elastic collisions, induced-radiations are separated into a diffusion-induced radiation term $C_{1\leftrightarrow 2}$ and a radiation term $C_{2\leftrightarrow3}$ induced by large-$q$ collisions with medium partons.
The formulae that follow immediately only applies to the Bethe-Heitler limit of incoherent radiation, where the formation time of the radiation $\tau_f$ is much shorter than the collisional mean-free-path $\elmpf$.
We will then introduce corrections to these formulae in the deep-Landau-Pomeranchuk-Migdal regime of $\tau_f\gg \elmpf$.

The large-$q$ collision-induced radiation term $C_{2\leftrightarrow3}$ is defined similarly as in equation \ref{eq:C22}, except that the radiation introduces another hard parton in the final state, labeled by four-momentum $k''$
\begin{eqnarray}
\nonumber
C_{2\leftrightarrow 3}^a(p_a) &=& \sum_{bcdef}\nu_b \int_{q_\perp>\Qcut} d\Pi^{p_b p_c}_{p_d p_e p_f} |M|^2_{p_b p_c;p_d p_e p_f}f^b(p_b)f_{0}(p_c)
\\\nonumber
&&
(2\pi)^3\left[
-\delta^{(3)}(p_a-p_b)\delta^{ab}
+\sum_{f=d,e,f}\delta^{(3)}(p_a-p_i)\delta^{ai}
\right]\\
&& + [\textrm{Absorption term}]
\label{eq:C23}
\end{eqnarray}
The first term on the right includes two-to-three-body collisions for parton radiation.
Its squared amplitudes are listed in reference \cite{Ke:2018jem} in the appendix.
The coupling at the radiation vertices is evaluated at scale $k_\perp$---the radiated parton's transverse momentum relative to the original parton.
Rigorously speaking, there are also detailed balancing processes corresponding to medium parton absorption.
However, the balancing term only becomes important when the distribution $f(p)$ is close to kinetic equilibrium.
Because we only consider particles with $p\cdot u >4T$ in the transport equation, the absorption term is neglected.

The diffusion-induced radiation collision integral is obtained by taking the $|q|\ll k_\perp$ limit of two-to-three-body squared amplitudes and integrating the phase-space over $|q|<\Qcut$.
In such an approximation, the collision integral reduces to an effective one-to-two-body integral, which is proportional to the soft transport parameter $\qhat_S$ in equation \ref{eq:qhat_soft},
\begin{eqnarray}
\nonumber
{\mathcal{C}}_{1\leftrightarrow 2}^a(p) &=& \sum_{bcd} \int dx d\bfk_\perp^2 \frac{\alpha_s(k_\perp)P_{cd}^b(x)}{(2\pi)(\bfk_\perp^2 + m_\infty^2)^2}  \\ && \left[-\qhat_{g,s}(p)f^b(p)\delta^{ab} + \frac{\qhat_{g,s}\left(\frac{p}{1-x}\right)f^b\left(\frac{p}{1-x}\right)}{1-x} \delta^{ac}
+ \frac{\qhat_{g,s}\left(\frac{p}{x}\right)f^b\left(\frac{p}{x}\right)}{x} \delta^{ad}
\right],
\label{eq:C12}
\end{eqnarray}
where $P_{cd}^b(x)$ is the vacuum splitting function of parton $b$ to parton $c$ and $d$ and we does not write down the explicit temperature dependence of the jet transport parameters.

\paragraph{Radiations in the deep-LPM region}
The diffusion- and collision-induced radiations discussed above only apply to those parton splittings whose formation time ($\tau_f$) is short compared to the collisional mean-free-path ($\elmpf$). 
However, collinear splittings in a dense medium often have $\tau_f \gg \elmpf$ in the so-called deep-LPM region.
We have to correct the above formulae to apply the transport equation to these splittings in the deep-LPM region. 
In reference \cite{Ke:2018jem}, we developed a simple correction method inspired by the next-to-leading-log induced radiation rate in \cite{Arnold:2008zu}.
First, one samples the incoherent radiation collision integral in equations \ref{eq:C23} and \ref{eq:C12} at time $t_1$.
Momentum broadening to the radiated partons and the formation time are determined self-consistently in the simulation,
\begin{eqnarray}
\tau_f &=& \frac{2x(1-x)E}{k_{\perp, t_2}^2},
\end{eqnarray}
where $k_{\perp, t_2}$ is the broadened transverse momentum at time $t_2 = t_1+\tau_f$.
Then, the radiation probability is suppressed by the factor
\begin{eqnarray}
P = \frac{\elmpfeff}{\tau_f} \sqrt{\frac{\ln(1+k_{\perp,t_2}^2/ m_D^2))}{\ln(1+\Qmax^2/m_D^2)}}.
\end{eqnarray}
$\elmpfeff= m_D^2/\qhat(E,T)$ is the effective mean-free-path.
$\Qmax$ is an estimation of the maximum momentum transfer in an elastic collision, which is taken to be the average center-of-mass energy between jet partons and medium partons $\Qmax = \sqrt{6ET}$.

This method has been tested in \cite{Ke:2018jem} and we refer to the detailed discussion there.
It was demonstrated to agree well with theoretical calculations in the deep-LPM region for induced radiations in an infinite static medium \cite{Arnold:2008zu}.
It also qualitatively captures the path-length \cite{CaronHuot:2010bp} and expansion rate dependence \cite{Baier:1998yf} of single radiation probability in a finite and expanding medium. 
To better demons rate that this model achieves quantitative agreement with our theoretical knowledge, we have included a detailed comparison between the theoretical single gluon emission rate and the {\tt LIDO} model simulations in a finite and expanding medium in appendix \ref{app:validate_lido}.

\begin{figure}
    \centering
    \includegraphics[width=.6\textwidth]{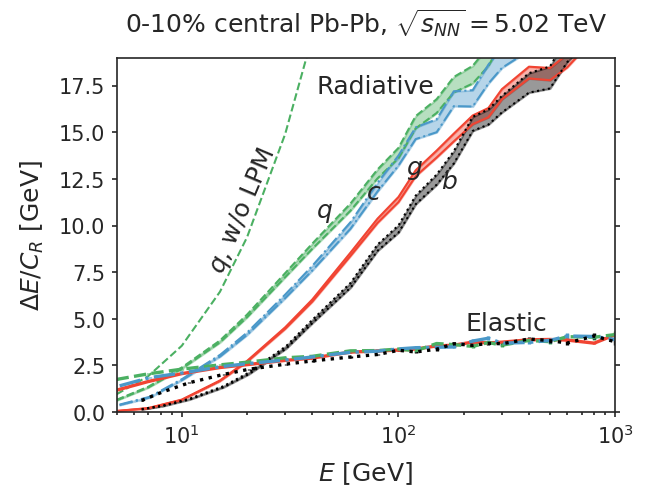}
    \caption{Parton energy loss in 0-10\% Pb-Pb collisions, scaled by the color factor, as a function of the initial parton energy. 
    The parton's initial production point is sampled from the transverse density of binary collisions.
    The radiative energy loss for light quark $q$, charm quark $c$, bottom quark $b$, and gluon $g$ are the colored bands labeled in the plot. 
    Lines corresponds to the elastic energy loss of different species of partons.}
    \label{fig:eloss}
\end{figure}

To demonstrate the importance of the LPM effect and provide a benchmark result for this transport model, we plot the simulated single-parton energy loss in 0-10\% Pb-Pb collisions in figure \ref{fig:eloss}.
Initial partons with energy $E$ at midrapidity with randomized azimuthal orientation.
The transverse positions are sampled according to the transverse density of binary nucleon collisions.
The energy loss is defined by the difference between the initial-state and final-state energy of the parton.
We show the elastic (lines) and radiative (bands) energy loss of gluons (red solid), light quarks (green dashed), and heavy quarks including charm (blue dash-dotted) and bottom (black dotted).
The parton energy loss has been rescaled by the color factor $C_R$ so the results are comparable for quarks and gluons.
In a finite and expanding medium, the elastic energy loss is a slowly varying function of parton energy and the radiative energy loss grows as $\ln E$ at high energy.
Elastic parton energy loss dominates at low energy $E<10$--$20$ GeV, while radiative energy loss dominates at high energy.
Moreover, the radiative parton energy loss of quark shows a clear mass dependence $\Delta E_q > \Delta E_c > \Delta E_b$.
Finally, we also include a calculation for light quark energy loss without the LPM effect (green dashed line) for comparison.
Without the LPM effect, radiative energy loss would increase faster than $\ln(E)$ at high energy.

\subsection{Medium excitations induced by hard partons: an approximate treatment}
\label{sec:model:response}
The linearized partonic transport model only handles hard particles with $p\cdot u>4T$.
The energy-momentum transfer to the soft sector ($p\cdot u<4T$) is considered in this section, and its impact on the final observables--transverse-momentum density $dp_T/d\eta/d\phi$--is modeled in a hydrodynamic-motivated ansatz.
This simple model consists of three parts: instant thermalization of energy-momentum transfer, angular redistribution of energy-momentum, and freeze-out.
\paragraph{Instant thermalization} We assume that, at each instant, the medium only responses to the four-momentum exchanges with jet partons\footnote{This means one neglects the perturbation to higher-order moments of the medium distribution function.}.
Within each time step $\Delta\tau$ and for each collision labeled by ``i'' at coordinate $(\tau_i, \bfx_{\perp,i} , \eta_{s,i})$, 
we compute the differences of the total four-momentum of hard particles in the initial state (IS) and in the final state (FS),
\begin{eqnarray}
\Delta p^\mu_i= \sum_{p\cdot u>4T}^{p\in \textrm{IS}} p^\mu - \sum_{p\cdot u>4T}^{p\in \textrm{FS}} p^\mu.
\label{eq:sources}
\end{eqnarray}
This gives the four-momentum deposited to the soft medium.

\paragraph{Angular redistribution of the energy-momentum perturbation}
\label{sec:response}
If $\Delta p^\mu_i$ only causes a small perturbation to the hydrodynamic evolution of the bulk medium, we can compute the linearized response of this energy-momentum deposition in energy-momentum density.
However, even the linearized response on top of an arbitrary hydrodynamic background is highly non-trivial.
We reduce the problem to a manageable level using the following simplifications:
\begin{enumerate}
\item Assume the typical temporal and spatial length scales of the perturbation are much shorter than those of space-time evolution of the background, including that of the transverse flow velocity, are neglected
\item The perturbations are localized in the space-time rapidity.
\item Neglect viscous effect.
\item Assume a constant speed-of-sound $c_s$.
\end{enumerate}
Under such assumptions, energy-momentum perturbations $e$ and $\mathbf{g}$ satisfy simple linearized ideal hydrodynamics in a static medium in the Bjorken frame.
Take one of the sources in equation \ref{eq:sources} and make the space-time Fourier transformation from $\tau, \bfx$ to  $\omega, \bfk$.
The resultant equations are\footnote{The equation for the shear mode $(1-\hat{\bfk}\hat{\bfk})\mathbf{g}$ is non-propagating and is omitted here and in the solution.},
\begin{eqnarray}
-i\omega e + i k \mathbf{g}\cdot\hat{\bfk} &=& \Delta p^0/\Delta \tau,\\
-i\omega \mathbf{g}\cdot\hat{\bfk}  + i c_s^2 k e &=& \Delta \mathbf{p}_i\cdot \hat{\bfk}/\Delta \tau,
\end{eqnarray}
where $k$ and $\hat{\bfk}$ are the magnitude and the unit direction vector of $\bfk$ and $\Delta p^0$ and $\Delta \mathbf{p}_i$ are the energy-momentum deposition.
The solutions are
\begin{eqnarray}
\label{eq:response_e}
e &=& \frac{1}{\Delta t}\frac{ \Delta p^0 \omega/k +  \Delta\mathbf{ p}\cdot\hat{\bfk}}{\omega^2-c_s^2k^2}k.\\
\mathbf{g} &=& \frac{1}{\Delta t}\frac{ \Delta p^0 c_s^2 k/\omega +  \Delta\mathbf{ p}\cdot\hat{\bfk}}{\omega^2-c_s^2k^2}\bfhatk.
\label{eq:response_g}
\end{eqnarray}
Now, we can avoid the transformation back to space-time coordinates by focusing on the angular distribution of energy ($d G^0/d\Omega$) and momentum ($d\mathbf{G}/d\Omega$)that are observed at a point far from the source.
Such distributions share the same angular structure as the numerators of equations
\ref{eq:response_e} and \ref{eq:response_g}; therefore,
\begin{eqnarray}
\label{eq:dedk-dpdk}
\frac{dG^\mu}{d\bfhatk}\equiv
\frac{d}{d\bfhatk}
\begin{bmatrix}
G^0\\
\mathbf{G}
\end{bmatrix}
 &=& \begin{bmatrix}
 1/c_s\\
 3\hat{\bfk}
 \end{bmatrix}\frac{\Delta p^0 c_s + \Delta\mathbf{p}\cdot \hat{\bfk}}{4\pi},
\end{eqnarray}
where $G^\mu = \int dr g^\mu$.
One can directly check the conservation of energy and momentum by integrating \ref{eq:dedk-dpdk} over $\hat{\bfk}$ and verify that  $\int dG^0  = \Delta p^0$ and $\int d\mathbf{G}= \Delta \mathbf{p}$.

The angular structure in equation \ref{eq:dedk-dpdk} is comprised of an isotropic term induced by energy deposition and a $\cos\phi_{\mathbf{p},\hat{\bfk}}$ modulated term induced by momentum deposition. 
Taking a light-like energy-momentum deposition $\Delta p^0=|\Delta \mathbf{p}|$ and choosing typical $c_s^2 \approx 0.25$, this ansatz propagates an excess of energy-momentum to the direction parallel to $\mathbf{p}$, but it propagates a small energy-momentum deficit to the direction anti-parallel to $\mathbf{p}$.
This qualitative feature is similar to those observed in the coupled transport-hydrodynamic simulation \cite{Chen:2017zte}.

\paragraph{Na\"ive freeze-out}
The formula in equation \ref{eq:dedk-dpdk} distributes energy-momentum as a function of the spatial solid angle $\hat{k}$. To calculate the contribution of perturbed transverse momentum density to jet energy as a function of azimuthal angle $\phi$ and pseudorapidity $\eta$, we convolve $dG^\mu/d\bfhatk$ with momentum-space information using a simple freeze-out procedure.

Though transverse flow is neglected in deriving spatial equation \ref{eq:dedk-dpdk}, flow effects are essential for particle distributions in the momentum space.
We assume, in central nuclear collisions, an averaged fluid velocity of $v_r = 0.6c$ at freeze-out and approximate the direction of the flow by $\hat{\bfk}$, so $u^\mu \approx (\gamma_\perp\cosh\eta_s, \gamma_\perp v_r\bfhatk_{\perp}, \gamma_\perp \sinh\eta_s)$.
Then, we apply the Cooper-Frye formula \cite{PhysRevD.10.186} on a hypothetical freeze-out surface $d\Sigma^\mu = d V u^\mu$
\begin{eqnarray}
\frac{d\Delta p_T}{p_T dp_T  d\eta d\phi} =  \int p_T (e^{-p\cdot (u+\delta u)/(T+\delta T)} - e^{-p\cdot u/T}) p\cdot d\Sigma,
\end{eqnarray}
where $\delta T = c_s^2\delta e/w$ and $\delta u^\mu =\delta g_\nu\cdot (g^{\mu\nu}-u^\mu u^\nu)/w$, where $w$ is the specific enthalpy of the background medium.
We approximate the system at freeze-out by a non-interacting gas of massless particles so that $w = 4T^4/\pi^2 $ per degree-of-freedom.
Putting these pieces together and expanding the result to the first order in $\delta T$ and $\delta u$, we arrive at the perturbation to the transverse momentum density
\begin{eqnarray}
\frac{d\Delta p_T}{ d\eta d\phi} &=&  \int p_T dp_T (p\cdot u) e^{-p\cdot u/T} \left(-\frac{p\cdot (1-u\cdot u)}{T}\cdot \frac{\delta g dV}{w}+\frac{p\cdot u}{T}\frac{c_s^2\delta e dV}{w}\right)\\
&=& \int p_T dp_T \frac{T e^{-p\cdot u/T}}{w} \left[\left(\frac{p\cdot u}{T}\right)^2(c_s^2+1) u_\mu - \frac{p\cdot u}{T}p_\mu\right] \frac{d G_{\bfhatk}^\mu}{d\bfhatk} d\bfhatk,\\
&=& \int\frac{3}{4\pi}\frac{(1+c_s^2)\sigma u_\mu-n_\mu}{\sigma^4}  \frac{d G_{\bfhatk}^\mu}{d\bfhatk} d\bfhatk,
\end{eqnarray}
where $\sigma = \gamma_\perp\left[\cosh(y-\eta_s)-v_\perp\cos(\phi-\phi_{\bfhatk})\right]$.
We remark that details of the medium, such as the number of species and degeneracy factors,  cancel in the final expression.
This is because we are computing the transverse-momentum density, which is less sensitive to the composition of the gas.
On the contrary, corrections to multiplicity distribution are sensitive to the chemical composition of the medium. 
Nevertheless, we will use this model to estimate corrections to charged-particle multiplicity in section \ref{sec:prediction:frag} when computing jet fragmentation function in nuclear collisions.

\subsection{Evolution outside of the medium, hadronization, and jet defintion}
\label{sec:model:vac2}

We assume the jet-medium interaction should become negligible when temperature drops to $T_f$ that is around or below the pseudo-critical temperature $T_c= 154$ MeV \cite{Bazavov:2014pvz}.
We call $T_f$ the termination temperature of jet-medium interaction.
For partons entering regions with $T<T_f$, they stop interaction with the medium according to the transport equation and we again use {\tt Pyhtia8} to evolve the parton shower in the vacuum.
The parton shower hadronizes through the Lund string fragmentation mechanism and is followed by hadronic decays.

The Lund string model requires the system consists of only a color singlet string of partons.
Colors of hard partons in the transport equation are initialized by the color generated by {\tt Pythia8} and the system, including beam remnants, is color neutral.
However, after the interactions with the QGP medium, the hard parton shower is not color-neutral in general.
Currently, it is intractable to trace the color exchange and conservation between hard partons and the entire medium.
This is partly because the derivation of the in-medium scattering cross-sections already takes the average over colors of both hard partons and the medium.
More importantly, even if we can keep the color information in the matrix-element of hard-soft collisions, it is impossible to model the transport of the soft carries of color charges in a strongly-coupled QGP.
In the present study, we use the following prescription to model color exchange between hard parton and the medium and impose color neutrality before applying the Lund string fragmentation model.
During the evolution, we assume the color of a hard parton is not altered in the small-angle diffusion process but is changed instantaneously in large-angle elastic collisions with a randomly sampled color exchange with the medium.
At the end of the transport evolution in QGP, we connect all possible final-state hard partons into strings.
The endpoints of some strings may carry net color charges due to interactions with the medium.
Then, thermal quarks or anti-quarks with the desired color or anti-color are sampled and attached to these endpoints so that the entire string is color neutral.
This procedure is similar to the one used in a study with the MARTINI model \cite{Schenke:2009gb}.
To guarantee energy-momentum conservation, the four-momentum of the sampled thermal partons will be treated as a negative source term to the medium using the ansatz for the hydrodynamic-like medium response that is introduced in section \ref{sec:model:response}.
Finally, each hard parton is assigned a virtually using the momentum broadening accumulated during the formation time of the latest medium-induced radiation $k_{\perp,\textrm{final}}$.
{\tt Pythia8} generates vacuum-like emissions out of the hot QGP medium from $k_{\perp,\textrm{final}}$ to the non-perturbative scale and performs hadronization.

We define jet at the hadronic level.
In each event, a grid of energy towers $E_{T, ij}$ is defined using the hadronic final state from hard partons and the contribution from linearized hydrodynamic response to transverse-momentum distribution,
\begin{eqnarray}
\nonumber 
E_{T, ij} &=& \frac{d \Delta p_T}{d\phi dy}(y_i, \phi_j) \Delta \phi\Delta y + \sum_{ 
\begin{aligned}
&{\scriptstyle |y_k-y_i|<\Delta y/2}\\[-10pt]
& {\scriptstyle |\phi_k-\phi_j|<\Delta \phi/2}
\end{aligned}}^N 
p_{T,k}.
\end{eqnarray}
Jets are defined from these ``energy towers'' using the anti-$k_T$ jet finding algorithm implemented in the {\tt FastJet} package~\cite{Cacciari:2008gp,Cacciari:2011ma}.
Because contributions to $E_{T,ij}$ from the linearized hydrodynamic response can be either positive or negative, in the first round of jet finding, we use only positive towers ($E_{T, ij}>0$) to define jets.
Then, the jet momentum is recalculated using all $E_{T, ij}$ contributions within the jet cone.
This method implicitly subtracts the unperturbed medium as the background from the jet energy.
Finally, Grid sizes $\Delta \phi$ and $\Delta \eta$ should be small compared to jet radii, but not too fine that increases jet clustering complexity significantly. 
In practice, we used a 300 by 300 grid within the acceptance $-3<y<3, -\pi<\phi<\pi$ with resolution $\Delta y = 0.020$ and $\Delta \phi=0.021$, respectively.

\section{Model calibration}
\label{sec:bayes}
The model introduced in the last section has multiple parameters and its predictive power relies on the ability to constrain these parameters from limited data.
In this section, we calibrate a subset of model parameters on the suppression of inclusive jet and hadron spectra data measured at the RHIC and the LHC using Bayesian inference techniques \cite{Bernhard:2016tnd}.
We will show that with reasonable choices of parameters, the model consistently describes the inclusive jet and high-$p_T$ hadron suppression within $\pm 10\%$-$20\%$ uncertainty.
More importantly, we report sets of representative parameters for predictions with uncertainty quantification in the next section.

\subsection{Summary of parameters}
\label{sec:bayes:parameter}
As the first work applying this model to jet quenching phenomenology, we only vary a subset of all possible parameters in the model.
In particular, the jet-medium interaction is completely determined by the perturbative jet-medium coupling at scale $\mumin$.
Next, we list the parameters that are tuned in the calibration with ranges of variation and explain why we fixed other parameters in this analysis.
\begin{enumerate}
    \item Parameter $\mumin$ in the jet-medium coupling, chosen to be proportional to the local QGP temperature. It serves as a cut-off for the running scale at finite temperature $\alpha_s(Q) = \alpha_s(\max\{Q, \mumin\})$. 
    The actual parameter varied is $\ln(\mumin/\pi T)$, ranging from $0.7$ to $4.0$. The appearance of $\pi$ is not necessary and one can simply translate it into $2.2 T < \mumin < 12.6 T$, which is around the typical energy of a thermal parton.
    \item The scale parameter $Q_0$ at which one transits jet shower evolution from high-virtuality vacuum parton shower evolution to transport evolution in the QGP. 
    $Q_0$ is assumed to only depend on the collision system, beam energy, and centrality.
    Practically, this is equivalent to two independent $Q_0$ parameters for central Au-Au collisions at the RHIC and central Pb-Pb collisions at the LHC respectively: $\QRHIC$ and $\QLHC$. 
    A current limitation is that the {\tt Pythia8} parameter that controls $Q_0$ is only allowed to be varied from 0.4 GeV to 2.0 GeV, which precludes the exploration of higher $Q_0$ values.
    We will consider other event generators in the future that allows a large range of variation.
    \item The termination temperature $T_f$ of QGP effects. Near the pseudo-critical temperature, the QCD equation of state deviates significantly from the Stefan-Boltzmann limit.
    To account for this gradual onset of confinement, we stop the transport evolution in hot QGP at $T_f$ that can be varied from 0.15 GeV to 0.17 GeV.
    Because the temperature drops slowly at late times of the hydrodynamic expansion, a $0.02$ GeV change in $T_f$ causes a large variation in the time duration that the system stays in the transport stage.
    This is demonstrated in appendix \ref{app:medium}.
\end{enumerate}
It is necessary to explain why we {\bf do not} vary the following parameters in this work.
\begin{enumerate}
    \item We do not vary the $\Qcut$ parameter introduced in section \ref{sec:model:lido}. Because given the perturbative input to the diffusion coefficients and large-$q$ collision matrix-elements, the  $\Qcut$ dependence of elastic energy loss due to the diffusion process is canceled by that of the large-$q$ collisions.
    This parameter is of interest if one intends to compare detailed effects due to diffusion dynamics versus large-$q$ collisions dynamics.
    The default choice in this study is $\Qcut=2m_D$.
    \item We do not vary ``hard'' particle cut-off whose default value is set at $4T$. 
    This parameter does not affect the energy loss of hard partons and therefore, high-$p_T$ inclusive hadron $R_{AA}$ spectra.
    Also, because both elastic recoil and hydrodynamic response transport energy to large angles, the jet suppression with $R=0.4$, for which we calibrated the model, is also insensitive to its choices.
\end{enumerate}

\begin{figure}
    \centering
    \includegraphics[width=.9\textwidth]{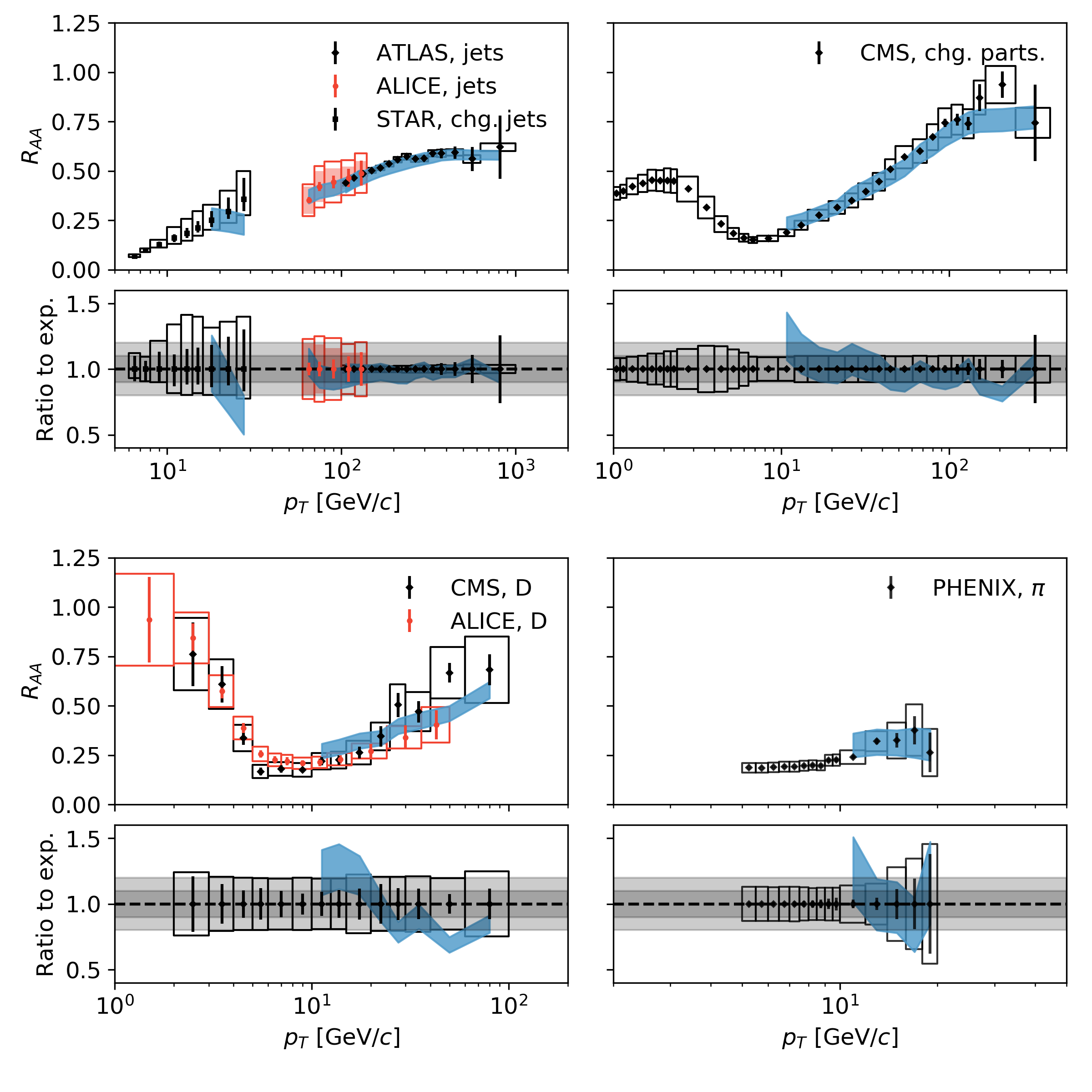}
    \caption{Emulated calculations of inclusive jet and single hadron suppression after model parameters are systematically calibrated to these observables. In each set of plots, the $R_{AA}$ is shown in the upper panel, and the ratio to experiment is shown in the lower panel. The blue bands correspond to 95\% credible limit given by the model emulator. The gray bands in the ratio plot denote $\pm 10\%$ and $\pm 20\%$ level of discrepancy respectively.}
    \label{fig:model2data}
\end{figure}

\subsection{Experimental dataset}
\label{sec:bayes:exp}
We calibrate parameters to inclusive jet and hadron suppression in central nuclear collisions.
Jet data include $R=0.4$ jet $R_{AA}$ in 0-10\% central Pb-Pb collisions at $\sqrtsNN=5.02$ TeV measured by the ATLAS Collaboration \cite{Aaboud:2018twu} and by the ALICE Collaboration \cite{Acharya:2019jyg}, $R=0.4$ charged jet $R_{AA}$ measured by the STAR Collaboration \cite{Adam:2020wen} in 0-10\% Au-Au collisions at $\sqrtsNN=200$ GeV. There is also a recent CMS measurement at this beam energy \cite{CMS-PAS-HIN-18-014}, but we do not include it in the calibration because its $p_T$ coverage is similar to those highest $\pTjet$ ATLAS data points.
We will compare the prediction to the CMS measurement after the calibration in section \ref{sec:prediction:jetR-jetshape}, especially focusing on the jet cone-size dependence.
The $\pTjet$ coverage at LHC ranges from 60 GeV to 1 TeV, while STAR data at RHIC extend the reach of $\pTchgjet$ down to 15 \GeVc\footnote{Note that the STAR Collaboration measurements charged jet, and uses a leading $\pTchg> 5$ GeV trigger in nuclear collisions. 
The triggering effect can bias the low-$p_T$ jet suppression. The STAR Collaboration has estimated the triggering bias to be small for $\pTchgjet>15$ \GeVc for $R=0.4$ jets.}.
Data on single inclusive hadron spectra include CMS measurement of charged particle $R_{AA}$ \cite{Khachatryan:2016odn} and D${}^0$ meson $R_{AA}$ \cite{Sirunyan:2017xss} in 0-10\% Pb-Pb collisions at $\sqrtsNN=5.02$ TeV and PHENIX measurement of pion $R_{AA}$ \cite{Adare:2012wg} in 0-10\% Au-Au collisions at $\sqrtsNN=200$ GeV.
We include $D$ meson data to make a connection to an earlier study of open heavy-flavor using the LIDO transport model

Only data on hadron $R_{AA}$ at $p_T>10$ \GeVc are used in the calibration for two reasons.
First, the soft-diffusion coefficient and the approximate scattering cross-sections for single-parton evolution are under better theoretical control in the high energy limit. 
Therefore, we limit our comparison to the single-hadron production at high $p_T$.
Second, the model only implements the Lund string hadronization but neglects the processes of parton recombination in the medium. 
The latter is important for hadron production at low and  intermediate $p_T$ ($p_T\lesssim 8$ \GeVc) \cite{Fries:2003vb}. 
The choice of $p_T>10$ \GeVc does have a certain degree of arbitrariness. 
In appendix \ref{app:bayes:pTcut} we checked the effect using different $p_T$ cut of 5 GeV$c$ and 15 \GeVc to see its impact on the parameter extraction.

\subsection{Parameter calibration}
\label{sec:bayes:posterior}
In this section, we focus on the results of the model calibration and provide detailed procedures of the Bayes parameter calibration in appendix \ref{app:bayes}.
In figure \ref{fig:model2data}, we demonstrate the level of agreement of the model (or model emulator) with data, after systematic parameter calibration.
Within each of the four plots in figure \ref{fig:model2data}, the upper panel shows the direct comparison to the data used for calibration and the lower panel shows the model-to-data ratios.
At the LHC energy, the calibrated model describes jet $R_{AA}$ within $\pm 10\%$ and charged particle suppression at high $p_T$ within $\pm 20\%$.
The shape of the calibrated D-meson $R_{AA}$ is slightly different from the CMS measurement and is closer to the ALICE measurement \cite{Acharya:2018hre}, though the latter is not included in the calibration.
At RHIC energy, the jet and pion $R_{AA}$ are consistent with data within experimental uncertainty.
The jet $R_{AA}$ at RHIC in the considered $p_T$ range decreases with $p_T$ while the STAR measurement increases. 

\begin{figure}
    \centering
    \includegraphics[width=.7\textwidth]{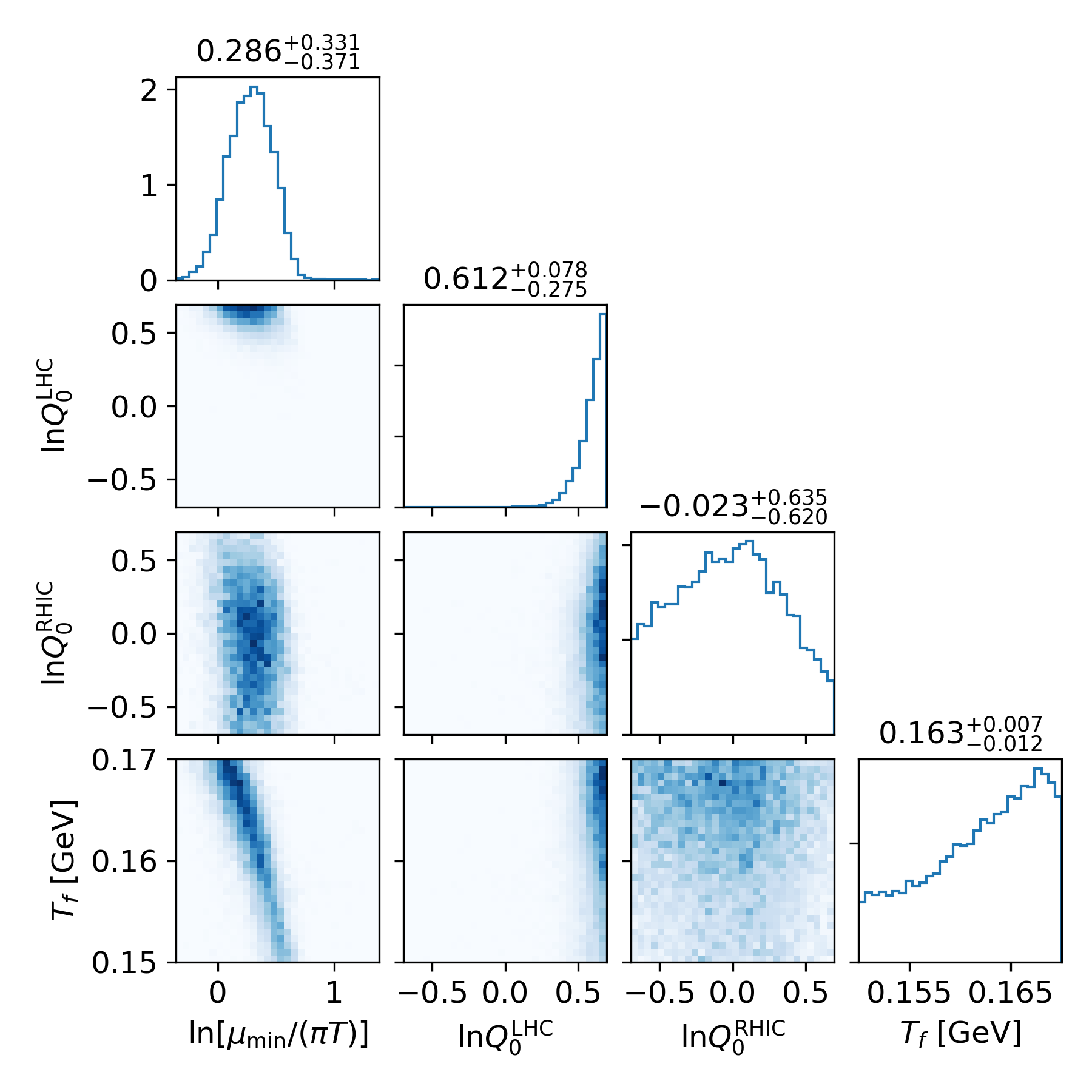}
    \label{fig:parameters}
    \caption{The posterior probability distribution of model parameters. Each diagonal plot shows single-parameter distribution after model-to-data comparison. Each off-diagonal plot corresponds to a two-parameter joint distribution. Numbers at the top of each diagonal plot are the median values with upper and lower uncertainties of each parameter, defined as the 95.7\% and 2.5\% quantile number of the one-parameter marginalized distribution.}
\end{figure}

The extracted probability distribution of model parameters (posterior distribution) is shown in figure \ref{fig:parameters}.
The logarithmic of $\mumin/\pi T$ has a well-constrained distribution around 0.3 with 95\% credible limits from -0.07 to 0.62. This range translates into a 95\% credible limits of $2.9 T< \mumin<5.8T$.
This is a reasonable range for the in-medium scale in the coupling constant at finite temperature, considering typical thermal momentum is about $3 T$.

The medium-scale parameter for parton shower at LHC $\QLHC$ is found to be large, with a 95\% credible limits from 1.39 GeV to 1.99 GeV. 
At the moment, we are not sure whether a larger $Q_0$ will be preferred if it is allowed to vary above $2$ GeV.
This scale parameter at RHIC energy $\QRHIC$ is not very well constrained: $1.0_{-0.47}^{+0.86}$ GeV.

\begin{figure}
    \centering
    \includegraphics[width=.9\textwidth]{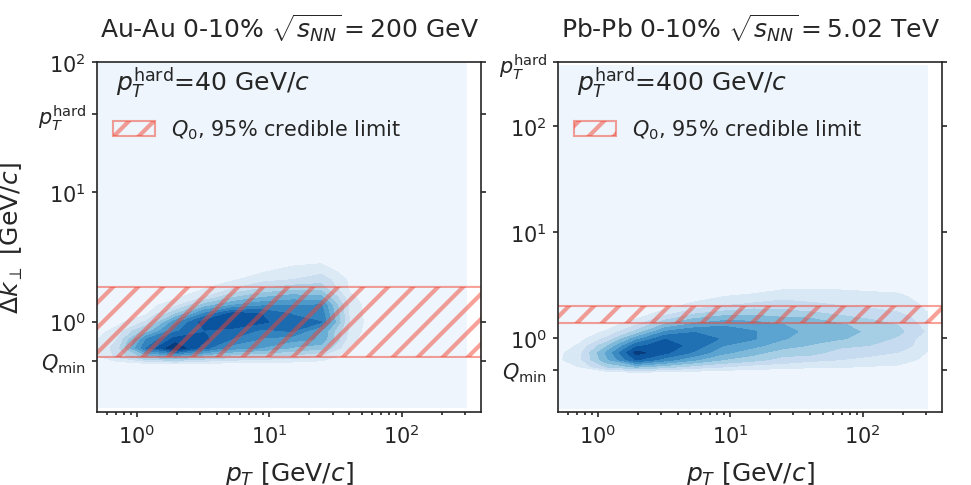}
    \caption{Comparing the calibrated $Q_0$ parameters to momentum broadening distributions obtained in simulations.}
    \label{fig:Q0_vs_kT}
\end{figure}

In section \ref{sec:model:vac}, we have argued that $Q_0$ should be comparable to the momentum broadening $\Delta k_T$ during the formation time of parton radiation.
In figure \ref{fig:Q0_vs_kT}, 95\% credible regions of $\QRHIC$ and $\QLHC$ (red shaded bands) are compared to the simulated distribution of $\Delta k_T$ as a function of the transverse momentum of the parton that splits.
Simulations use typical $\pThard=40$ \GeVc for $\sqrtsNN=200$ GeV and $\pThard=400$ \GeVc for $\sqrtsNN=5.02$ TeV.
First, we find that values of $\Delta k_T$ increase rather slowly with $p_T$ and occupy a relatively compact region as compared to the range of virtuality from $\pThard$ to $\Qmin= 0.5$ GeV during the evolution.
Both features are in favor of approximating the separation scale between vacuum and medium-induced shower by a single $Q_0$ value.
At the LHC energy, typical values of $\Delta k_\perp$ are slightly lower than $\QLHC$, but still around the same order of magnitude for partons with $p_T>10$ \GeVc.
At RHIC energy, the large uncertainty of $Q_0$ covers the same range of $\Delta k_\perp$ distribution.

Finally, the termination temperature of jet-medium interaction $T_f$ prefers large values towards 0.17 GeV, though it is in general not well constrained.
Note that $T_f$ strongly correlates with the in-medium scale parameter in the coupling constant as shown in the bottom-left corner plot:
a larger$\mumin$--smaller coupling strength-- is correlated with a smaller termination temperature, which means the jet-medium interaction lasts longer to compensate for the reduced coupling in the calibration.

\subsection{Parameter sets for central prediction and error estimation}
\label{sec:bayes:psets}
In principle, to make predictions with quantified uncertainty from the calibrated model, we should pull random parameter samples from the posterior distribution and make an ensemble of predictions using the sampled parameters; then, central values and uncertainties can be defined accordingly from the ensemble prediction.
However, it is unnecessary and {\bf impractical} to propagate arbitrary many parameter samples to predictions.
Instead, we define a small number of representative parameter sets to characterize parameter variations in the high-likelihood region of the posterior, including a central parameter set and a few error estimation parameters sets for the prediction.

\begin{table}[ht]
    \centering
    \begin{tabular}{c|cccc}
    \hline
    $\mathbf{p}$ & $\mumin$ & $\QLHC$ [GeV] & $\QRHIC$ [GeV] & $T_f$ [GeV] \\
    \hline
         Central set & 1.313 & 1.825 & 0.988 & 0.1621 \\
         Error set 1 & 0.989 & 1.944 & 1.057 & 0.1699 \\ 
         Error set 2 & 1.854 & 1.690 & 0.910 & 0.1526 \\ 
         Error set 3 & 1.307 & 2.000 & 1.527 & 0.1585 \\ 
         Error set 4 & 1.319 & 1.590 & 0.579 & 0.1664 \\ 
         Error set 5 & 1.329 & 2.000 & 0.689 & 0.1608 \\ 
         Error set 6 & 1.293 & 1.593 & 1.512 & 0.1636 \\ 
         Error set 7 & 1.459 & 1.864 & 1.030 & 0.1650 \\ 
         Error set 8 & 1.146 & 1.777 & 0.936 & 0.1583 \\ 
    \hline
    \end{tabular}
    \caption{Parameter sets for central prediction and error estimation. They are determined from the posterior probability distribution of parameters using a procedure described in the text.}
    \label{tab:sets}
\end{table}

\begin{figure}
    \centering
    \includegraphics[height=.41\textwidth]{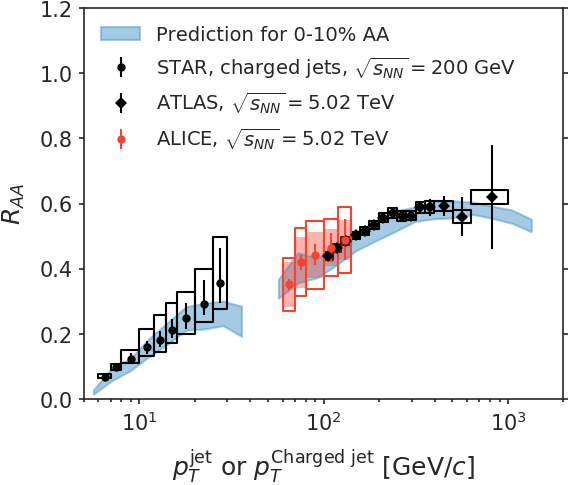}
    \includegraphics[height=.41\textwidth]{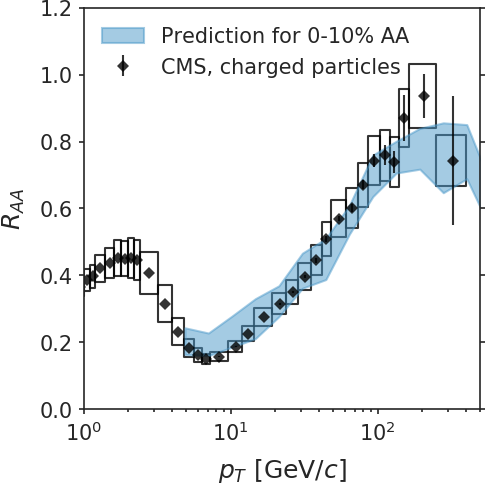}
    \caption{Validation of the representative parameter sets by performing full model calculations using the nine sets of parameters listed in table \ref{tab:sets}. The bands corresponds to the envelop of the nice predictions.
    }
    \label{fig:model-validation}
\end{figure}

There is not a unique way to define the representative parameter sets.
The method adopted here relies on the posterior showing a small degree of non-linear correlations among different parameters. 
Neglecting the non-linear correlations, we define linear combinations of parameters $q_i = c_{ij} p_j$  so that new parameters $q_i$ are linearly independent from each other.
Then in the $q$-space, the central parameter set is defined to be the collection of medians of parameters,
\begin{eqnarray}
\mathbf{q}_{\textrm{central}} = \{\mathrm{median}(q_0),\cdots, \mathrm{median}(q_3) \},
\end{eqnarray}
and error sets are obtained by changing one of the four values in the central set to the $2.5\%$ and $97.5\%$ quantile numbers of that parameter,
\begin{eqnarray}
\mathbf{q}_{\textrm{error},k\pm} = \{\mathrm{median}(q_0),\cdots, \mathrm{quantile}_{2.5\%}^{97.5\%}(q_k), \cdots \mathrm{median}(q_3) \},\quad k=0,1,2,3.
\end{eqnarray}
This introduces eight error parameter sets in total.
Finally, $\mathbf{q}_{\textrm{central}}$ and $\mathbf{q}_{\textrm{error},k\pm}$ are transformed into the original model parameter space $\mathbf{p}$ with values listed in table \ref{tab:sets}.

As a validation, we compute the single inclusive jet and hadron suppression using the representative inputs in table \ref{tab:sets} and show the results in figure \ref{fig:model-validation}.
Uncertainty bands are drawn as the envelope (blue bands) of these nine different calculations.
The uncertainty is at about $\pm15\%$ level, compatible with the emulator predictions that use random draws of samples from the full posterior distribution.
These representative parameter sets will be used for predictions and analysis for jet suppression and modification.

\section{Energy flow induced by jet-medium interaction}
\label{sec:energy-flow}
Before using the calibrated model to predict jet modifications in section \ref{sec:prediction}, we would like to understand how energy-momentum is transported within the parton shower.
We studied a simple case---back-to-back quark or gluon pairs produced with fixed $\pThard$, using the central parameter set in table \ref{tab:sets}.
We focus on the energy distribution $E(\zhard, r)$ and $E(p_T,r)$ of partons in the two-dimensional spaces ($\zhard$, $r$) and ($p_T$, $r$), where $\zhard$ is the longitudinal momentum fraction relative scale of the hard process $\pThard$ and $r$ is the radial distance of a particle relative to initial hard parton,
\begin{eqnarray}
E(\zhard, r) \equiv \zhard\frac{dN}{d\ln \zhard dr},
\end{eqnarray}
or in transverse momentum $p_T$ and $r$,
\begin{eqnarray}
E(p_T, r) \equiv  p_T\frac{dN}{d\ln p_T dr}.
\end{eqnarray}
The subscript ``H'' reminds us that the reference scale of $\zhard$ is the momentum of the initial hard process. 
The definition of $\zhard$ if different from that of $\zjet$ used for inclusive jet fragmentation function measurement, whose 
reference scale is the transverse momentum of the reconstructed jet.
We will comment on the difference between $\zhard$ and $\zjet$ at the end of this section and emphasize again in section \ref{sec:prediction:frag}.
This simple example elucidates the respective roles of elastic collisions, induced radiation, and medium excitations in modifying $E(\zhard, r)$ and $E(p_T,r)$, which helps to interpret the prediction in section \ref{sec:prediction}.

We initialize pairs of back-to-back quarks or gluons with initial hard momentum $\pThard$ at mid-rapidity.
Following the steps described in section \ref{sec:model}, we use \pythia to generate vacuum parton showers and initialize the transport equation in the medium at scale $Q_0$.
Toward the end of transport in the medium at $T_f$, the jet shower continues to evolve in vacuum down to $Q_0=0.5$ GeV.
We omit hadronization in this simple example.

\subsection{Energy transport via vacuum and induced radiations}
\label{sec:energy-flow:vac}
We first focus on energy transport induced by radiative process and momentum broadening due to elastic collisions, while elastic energy loss and medium excitation are turned off\footnote{This is achieved by rescaling the magnitude of a particle's momentum to its value before the elastic collision or a diffusion step.}.

Energy cascades due to radiative processes in the vacuum and in a static medium were studied in \cite{Blaizot:2013hx,Blaizot:2014ula,Fister:2014zxa,Blaizot:2014rla,Blaizot:2015jea} under certain approximations. 
Here is one of its key observations.
Compared to the vacuum-like jet shower governed by DGLAP-type evolution with vacuum splitting kernels $P^{(0)}(x)$, parton splittings in the medium has an enhanced soft splitting rate because its typical inverse formation time increases for softer parton energy.
For example, in a static medium, $dR/dx \sim \alpha_s P^{(0)}(x)\tau_f^{-1} \sim \alpha_s P^{(0)}(x) \sqrt{\qhat/xE}$, where $x$ is the momentum fraction of the softer daughter parton relative to the momentum of the mother parton.
Such behavior builds up a power-law like energy spectrum $dE/d\zhard \sim \zhard^{-1/2}$ for $\zhard \ll 1$, which causes a finite amount of energy to flow into $\zhard=0$ within any arbitrary time duration \cite{Blaizot:2015jea}.
In practice, the ``point-like'' energy sink at $\zhard=0$ is replaced by a finite region $z\lesssim z_{BH}$.
$z_{BH}$ is the energy fraction corresponding to the Bethe-Heitler energy scale $z_{BH} = \omega_{BH}/E$ with $\omega_{BH}\equiv \frac{1}{2} \qhat\elmpf^2$.
The flow of energy ceases below $z_{BH}$, because the Bethe-Heitler radiation rate $dR_{BH}/dx \sim \alpha_s P^{(0)}(x) \elmpf^{-1}$ lacks additional enhancement at small-$x$.
While splittings from higher $\zhard$ regions continue to pump energy into this region, energy starts to accumulate among particles with typical energy $\omega_{BH}$.

Though some detailed assumptions employed in \cite{Blaizot:2015jea} become invalid in a finite and expanding plasma, one still preserves the most important qualitative feature that the inverse formation time of induced radiations increases for soft splittings.
One also notices that as a consequence of the formation of the thermal medium such that $\qhat\sim \alpha_s^2 T^3$ and $\elmpf\sim 1/(\alpha_s T)$, the Bethe-Heitler energy $\omega_{BH}=\mathcal{O}(T)$ coincides with the thermal scale\footnote{Neglecting the logarithmic factor, the jet transport parameter for a gluon is $\hat{q} \approx \alpha_s C_A T m_D^2$. Effective mean-free-path has been defined as $m_D^2/\hat{q}$. These lead to $\omega_{BH} \approx \pi T$.
}.
This suggests that induced-radiation processes alone can build up energy storage near the thermal scale $p\sim T$, no matter how energetic the initial hard jets are.
Such behavior has also been confirmed in recent numerical simulations including the effective kinetic theory with resummed radiation vertex \cite{Schlichting:2020lef} and the {\tt LIDO} transport model in the large-medium $L_{\textrm{med}}\gg \tau_f$ limit, which is discussed in section 3.5.3 of \cite{Ke:2020coj}.
We will see immediately that this feature indeed emerges from our simulation.

Because $\omega_{BH}$ coincides with the thermal scale, the distribution of the accumulated energy-momentum around $\omega_{BH}$ will be further modified by other transport mechanisms including elastic collisions and jet-induced medium responses that are important for particles with $p\sim T$.
This will be discussed in section \ref{sec:energy-flow:full}.

\begin{figure}
    \centering
    \includegraphics[width=.75\textwidth]{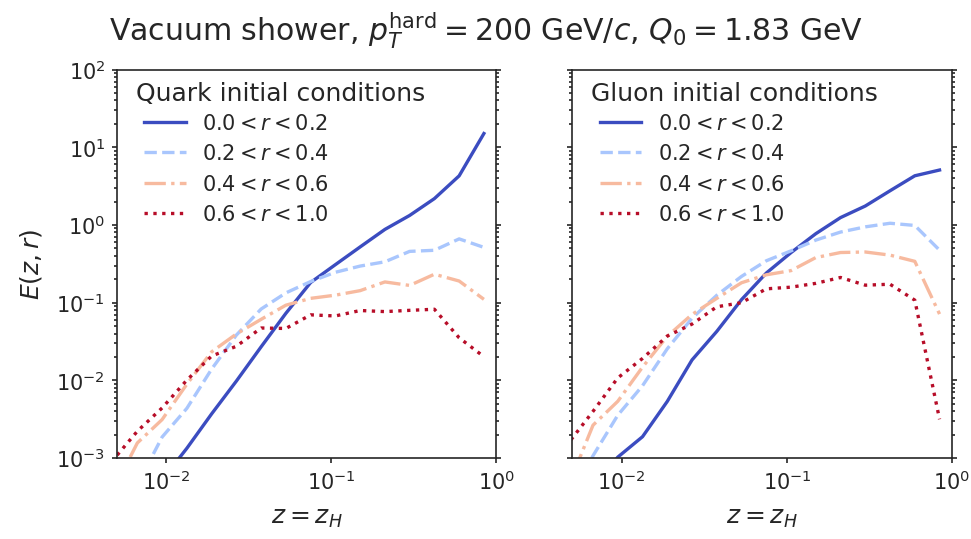}
    \caption{
    $E(\zhard,r)$ after vacuum parton shower evolution down to $Q_0$.
    }
    \label{fig:PsiR_vac}
\end{figure}
\begin{figure}
    \centering
    \includegraphics[width=.79\textwidth]{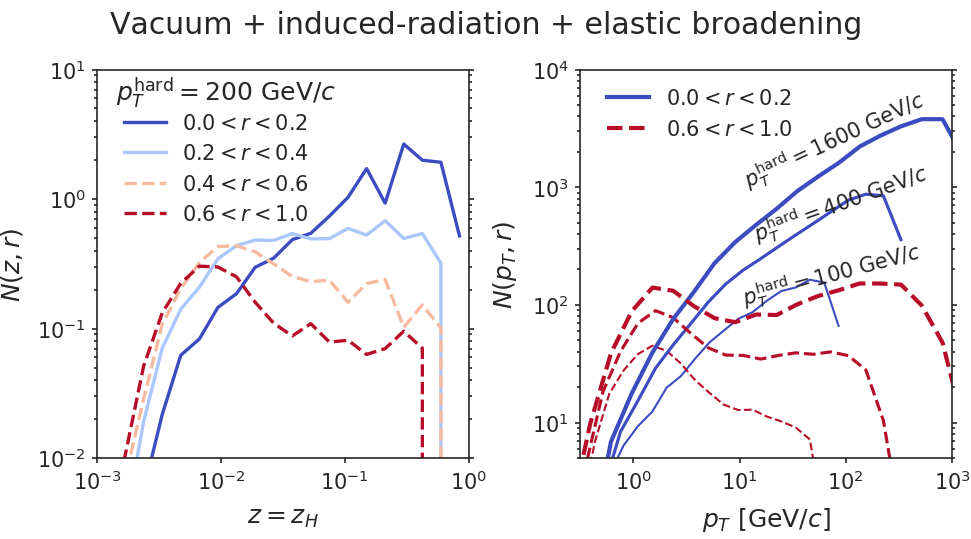}
    \caption{Energy distribution $E(\zhard,r)$ that includes effects from radiative process (vacuum plus medium-induced) and elastic broadening effects.
    Left: $E(\zhard,r)$ for $\pThard= 200$ \GeVc at various radii.
    Right: $E(p_T,r)$ at small and large radii for $\pThard=$100, 400, and 1600 \GeVc respectively.
    }
    \label{fig:PsiR_rad}
\end{figure}

In figure \ref{fig:PsiR_vac}, we plot the energy distribution $E(\zhard, r)$ from the vacuum-like parton shower evolution of a jet produced at $\pThard=200$ \GeVc to a final scale $Q_0=1.83$ GeV.
Results shown in the left and the right plots of figure \ref{fig:PsiR_vac} are for initial quarks and gluons, respectively.
Vacuum radiations transfer the energy of the initial hard quark/gluon into a broad range of $\zhard$ and $r$, while most of the energy remains in the small-$r$ and high-$\zhard$ region.
The distribution from an initial quark is harder and narrower than that from an initial gluon, as gluons radiate more frequently. 

Next, we evolve the jet shower in the QGP medium with induced radiations and the elastic broadening, following it by vacuum radiations outside the QGP.
The modified $E(\zhard,r)$ for a initial gluon is shown on the left panel of figure \ref{fig:PsiR_rad}.
Compared to figure \ref{fig:PsiR_vac} (right), the energy distribution is softened and widened.
More importantly, an ``energy bump'' build up in the large-$r$ and low-$\zhard$ (and small $p_T$) region.
To confirm that this bump is indeed the consequence of radiative energy transfer to the thermal scale, we show the $p_T$-dependent energy distributions $E(p_T, r)$ with different $\pThard$: 100, 400, and 1600 \GeVc on the right panel of figure \ref{fig:PsiR_rad}.
Evidently, changing $\pThard$ by an order of magnitude does not affect the typical momentum scale $p_T\sim 2$ \GeVc at which a bump develops.
Therefore the bump indeed locates at a soft medium scale that is independent of the initial hard scale.

\subsection{Energy transport with full effects: radiation, elastic energy loss, and medium excitation}
\label{sec:energy-flow:full}

\begin{figure}
    \centering
    \includegraphics[width=.8\textwidth]{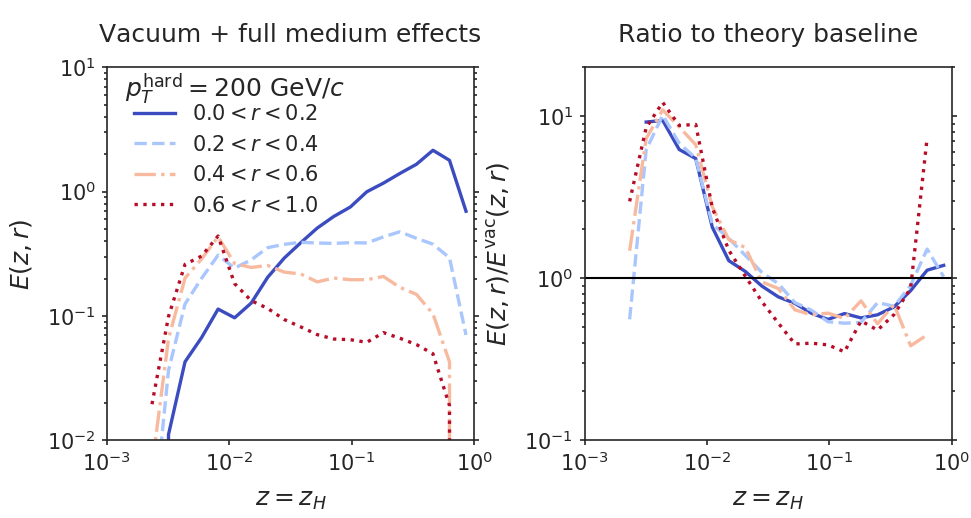}
    \caption{Energy distribution $E(\zhard,r)$ with full medium effects.
    Left: $E(\zhard,r)$ for $\pThard= 200$ \GeVc at various radii.
    Right: ratio between the energy distribution with and without medium effects $E(\zhard,r)/E^{\textrm{vac}}(\zhard,r)$ with $\pThard=200$ \GeVc.
    }
    \label{fig:PsiR_full}
\end{figure}

We have seen in figure \ref{fig:eloss} that the elastic energy loss is almost independent of the momentum of a parton.
In the meantime, the parton multiplicity at low-$\zhard$ is much higher than that at high-$\zhard$ during jet evolution in the QGP.
Combining these two arguments, elastic energy loss is most important for low-$\zhard$ region while becomes insignificant at high-$\zhard$.
The lost energy-momentum due to elastic scatterings are then redistributed by either hydrodynamic responses or hard medium recoiled partons.
The former induces near-equilibrium perturbation that modifies particle distribution around $p_T\sim T$ at large angles.
Momentum of the recoil particles in a $t$-channel process range from $gT$ to the average center-of-mass energy $\sqrt{6ET}$.
Therefore, the inclusion of elastic energy loss plus medium excitation mainly affects particles within $\mathcal{O}(T)<p_T<\mathcal{O}(\sqrt{ET})$.

On the left of figure \ref{fig:PsiR_full}, we show the resulting $E(\zhard,r)$ with full medium effects: induced-radiations, elastic broadening, and energy loss, and the medium excitation.
Comparing to the left of figure \ref{fig:PsiR_rad}, elastic energy loss plus medium excitation deform the low-$\zhard$ bump into a sharper peak around $p_T\sim 1$--2 \GeVc ($\zhard=0.005$--0.01).

On the right of figure \ref{fig:PsiR_full}, we show the ratio of the modified energy distribution to the theoretical baseline $E^{\textrm{vac}}(\zhard,r)$ due to parton shower in vacuum down to $Q_0=0.5$ GeV.
There is an energy deficiency at intermediate $\zhard$ compared to the theoretical baseline in the vacuum, and energy excess develops at the small $\zhard$ with soft momentum.
This qualitatively agrees with the finding of authors of \cite{PhysRevLett.120.232001,Caucal:2020xad}, where both vacuum and BDMPS-Z type medium-induced emissions are included.
However, we emphasize that in our simulation the energy excess is a combined result of both medium-induced radiation and jet-induced medium excitations:
induced radiation builds up the energy excess at the soft-scale and jet-induced medium excitation reshapes the bump into a sharper peak closer to the thermal momentum. 

When $\zhard$ approaches one, the ratio can be larger than unity.
This is a consequence of the way we interface the vacuum-like evolution equation and the transport equation and we explain as follows.
In our calculation for jets in medium, the scale evolution from $Q_0$ to $k_{\perp, \textrm{final}}$, where $k_{\perp, \textrm{final}}$ is the momentum broadening accumulated during the formation of the last medium-induced radiation, is replaced by the evolution in QGP using the transport equation.
In figure \ref{fig:eloss}, we have already seen that the high-energy parton energy loss in the medium grows logarithmically with parton energy: $\Delta E_{\textrm{med}} \sim \ln E$.
In simulations for jets the vacuum, soft vacuum radiation between $Q_0$ and $k_{\perp, \textrm{final}}$ causes energy shifts of the leading parton $\Delta E_{\textrm{vac}}(\frac{Q_0}{k_{\perp, \textrm{final}}})$.
Because $\ln\frac{Q_0}{k_{\perp, \textrm{final}}}\lesssim \ln\frac{Q_0}{\Qmin}\approx 1.2$ is not a large number, we can estimate $\Delta E_{\textrm{vac}}$ as  $\frac{\alpha_s}{2\pi} C_R \int_0^{1/2} xE P(x) dx \frac{dQ^2}{Q^2}\propto \alpha_s C_R E\ln\frac{Q_0^2}{k_{\perp, \textrm{final}}^2}$.
As a result, at large energy, the $\Delta E_{\textrm{vac}}$ that we removed from scale evolution is larger than the energy loss in the medium caused by jet-medium interactions, which explains why the ratio is larger than unity for $\zhard\sim 1$ for this simple test where $\zhard = \pTparton/\pThard$.
This phenomenon is also discussed in \cite{Caucal:2020xad}\footnote{In \cite{Caucal:2020xad}, $Q_0\sim\sqrt{\qhat E}$ is an energy-dependent scale estimated from multiple-collision effects in a static medium; while in our simulation, $Q_0$ is independent of energy as suggested from figure \ref{fig:Q0_vs_kT}.}. 

However, such an excess is only visible when we focus on very energetic particles so that $\Delta E_{\textrm{vac}} > \Delta E_{\textrm{med}}$ and analyze the result using the variable $\zhard = \pTparton/\pThard$.
In fact, the peak near $\zhard=1$ can be smaller than unity for a lower-$\pThard$ or be smeared out by hadronization effects.
Furthermore, in inclusive measurement, the momentum fraction is defined using $\zjet = p_T/\pTjet$. 
Unlike using $\pThard$, jets in a sample with a fixed $\pTjet$ in nuclear collisions starts are produced with a higher scale and favors jets with harder constituents due to jet energy loss.
Therefore, the jet fragmentation function ratio between AA and pp collisions being large than one at large $\zjet$ is mainly the result of the trigger bias \cite{Spousta:2015fca,Caucal:2020xad}. 
To remove the trigger bias and test this effect due to a switching scale between vacuum-like and medium-induced radiations, one would like to use $\gamma$-tagged and $Z$-tagged jets samples \cite{Chen:2020tbl}.
Using $z_{\gamma/Z}=p_T/p_T^\gamma$ or $p_T/p_T^Z$, one should focus on the ratio of fragmentation function large-$z_{\gamma/Z}$ and its dependence on the transverse momentum of the trigger-boson.
It would be interesting to see how this effect manifests in the fragmentation functions of $\gamma$-jet and $Z$-jet. 
It can also help to test the current modeling of the switching between vacuum-like evolution and in-medium evolution.
This will be pursued in future studies.

\subsection{Remarks on trigger bias}
The energy flow picture discussed in this section helps understand the theoretical origin of certain features that we are going to see in data.
But experimentally, it is impossible to determine the exact initial hard scale $\pThard$ for jet production in both nuclear collisions and proton-proton collisions.
Experiments usually compare single inclusive jets in nuclear and p-p collisions at the same final $\pTjet$ . 
One has to be careful when interpreting the difference between jet properties in nuclear and the p-p baseline.
For example, jets in central Pb-Pb collisions are produced at a higher scale on average than the reference jets in p-p collisions due to energy loss, which means ratios deviating from unity are not always direct consequent of energy flow.

\section{Prediction of jet modifications}
\label{sec:prediction}

In this section, we predict modifications to jet fragmentation function and jet shape, as well as the cone-size dependence of jet $R_{AA}$.
These predictions use full-model calculation with representative parameter sets from the calibration in section \ref{sec:bayes:posterior}.
Jets are defined at the hadron level with the same kinematics cuts in the measurements.

\subsection{Modified jet fragmentation function in central Pb-Pb collisions}
\label{sec:prediction:frag}

\begin{figure}
    \centering
    \includegraphics[width=.6\textwidth]{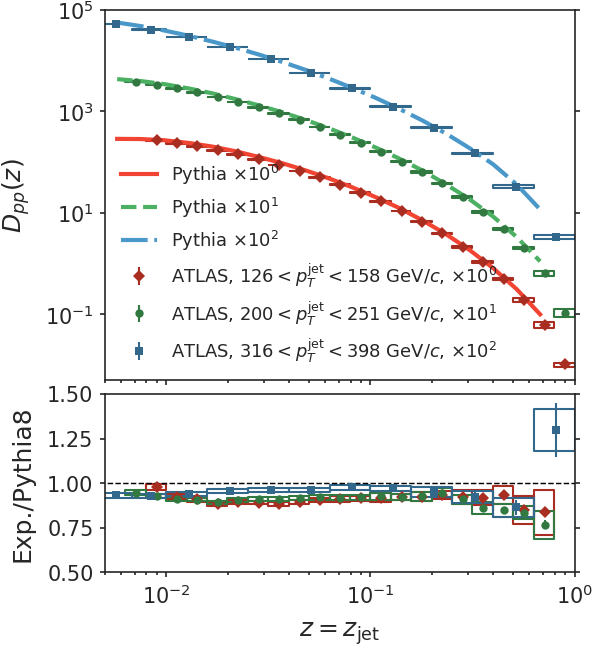}
    \caption{Comparison of charged particle fragmentation function of inclusive jets obtained by {\tt Pythia8} simulation to data from the ATLAS experiment \cite{Aaboud:2018hpb}. Ratio of experiment to simulation is shown in the lower panel.}
    \label{fig:Dz_baseline}
\end{figure}

For the ATLAS measurement \cite{Aaboud:2018hpb}, the jet fragmentation function is either defined as charged particle distributions in longitudinal momentum fraction $\zjet$ or in transverse momentum $p_T$, normalized by the number of jets,
\begin{eqnarray}
D(\zjet) &=& \frac{1}{N_{\textrm{jet}}}\frac{dN_{\textrm{ch}}}{d\zjet}, \quad \zjet = \frac{p^{\textrm{ch}}\cdot p^{\textrm{jet}}}{(\pTjet)^2} \approx \frac{\pTchg\cos r}{\pTjet},\\
D(p_T) &=& \frac{1}{N_{\textrm{jet}}}\frac{dN_{\textrm{ch}}}{dp_T}.
\end{eqnarray}
Note that $\zjet$ uses the transverse momentum of the reconstructed jet as the reference scale, which is different from that for $\zhard$ in section \ref{sec:energy-flow}.

In figure \ref{fig:Dz_baseline}, we compare the {\tt Pythia8} simulation as a baseline to the ATLAS measurement of $D(z)$ in p-p collisions~\cite{Aaboud:2018hpb} for $126<\pTjet<158$ \GeVc, $200<\pTjet<251$ \GeVc, and $316<\pTjet<398$ \GeVc.
There is a reasonable agreement for a wide range of $\zjet$ except in the very large-$\zjet$ bins, where the simulation drops faster than data toward $\zjet=1$.

\begin{figure}
    \centering
    \includegraphics[width=\textwidth]{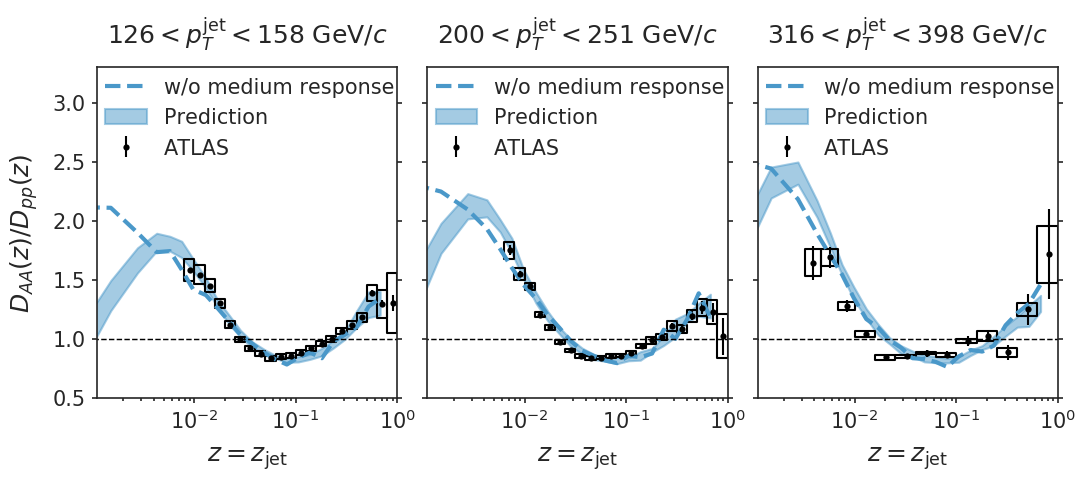}
    \caption{
    Prediction of modification to the jet fragmentation function ($D(z)$) and comparison to ATLAS measurements for three different $\pTjet$ bins \cite{Aaboud:2018hpb}. Bands corresponds to the envelop of calculations using the parameter sets listed in table \ref{tab:sets}. Dashed lines are calculations using the central parameter set but excluding contributions from medium excitations when calculating fragmentation functions.
    }
    \label{fig:frag_z}
\end{figure}
\begin{figure}
    \centering
    \includegraphics[width=.88\textwidth]{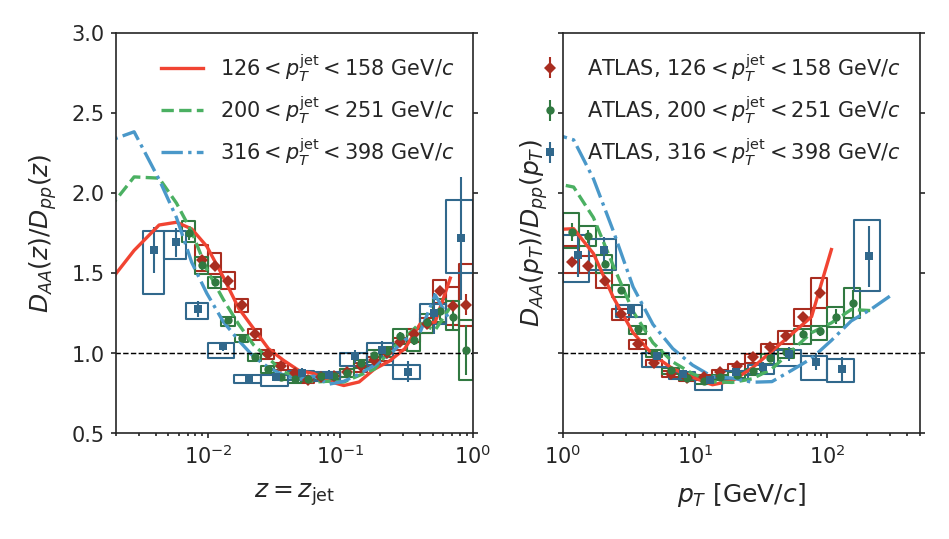}
    \caption{ An investigation of the scaling properties of the calculated jet fragmentation function in Pb-Pb collisions using the central prediction parameter set in table \ref{tab:sets}. Left: overlapping calculations and measurements of $D(z)$ ratios for three different $\pTjet$ bins. Right: Left: overlapping calculations and measurements of $D(p_T)$ ratios for three different $\pTjet$ bins.}
    \label{fig:frag_z_pT}
\end{figure}

Our predictions for ratios between jet fragmentation function $D(z)$ in Pb-Pb and p-p collisions are shown in figure \ref{fig:frag_z}.
The predictions include bands of uncertainty using the representative parameter sets in table \ref{tab:sets}.
The model constrained by inclusive jet and hadron suppression consistently describes the modification of jet fragmentation function observed in experimental data.
To illustrate the respective roles of radiative processes and medium excitations, we compare the full model calculation to the calculation that excludes contributions from medium excitations (dashed lines).
We emphasize that jets are still defined with contributions from medium excitations so that the jet samples are unbiased.
It is clear that even without contributions from medium excitations, low-$z$ enhancements of the ratio $D_{AA}(p)/D_{pp}(z)$ are already sizeable.
In the previous section, we have discussed that the medium-induced radiative process effectively transfers energy towards small-$z$ and causes the accumulation of energy at thermal scales.
Because the overall impact of jet-medium interactions is to excite medium partons to higher energy, contributions from medium excitations are negative and become positive at high transverse momenta. 
As a result, medium excitations develop a notable peak in the modification of fragmentation functions around $p_T\sim 0.7$ GeV for all three $\pTjet$ bins.

Modifications of $D(\zjet)$ are known to demonstrate different scaling behaviors in the hard and the soft regions \cite{Aaboud:2018hpb}:
ratios for $D(\zjet)$ at high-$\zjet$ for different $\pTjet$ collapse on the same curve; while ratios for $D(p_T)$ at low-$p_T$ are approximately independent of the $\pTjet$.
In figure \ref{fig:frag_z_pT}, the central predictions for $D(z)$ (left) and $D(p_T)$ (right) ratios for three different jet $p_T$ bins are compared with experimental data.
The high-$\zjet$ scaling of $D(z)$ is well reproduced.
In the model, the energies of high-$\zjet$ particles are sufficiently higher than other scales such as temperature and elastic recoil energy from  $\mathcal{O}(T)$ to $\mathcal{O}(\sqrt{ET})$; therefore, it is natural that its modification, dominated by induced radiation, is solely a function of $\zjet$.
For the modifications of $D(p_T)$, the model captures the low-$p_T$ enhancement of $D(p_T)$ at $p_T\lesssim 4$ \GeVc.
From discussions in section \ref{sec:energy-flow}, we see that the origin of low-$p_T$ scaling is a combination of both induced-radiation and medium excitation: energy first builds up at soft-momentum scale due to induced radiation, whose distribution is further modified by the jet-induced medium response.
Comparing the magnitude of the enhancement for different $\pTjet$, the $\pTjet$ dependence in the simulation is stronger than that observed in experimental data.
This could be caused by several reasons.
For example, it is possible that the transition scale parameter $Q_0$ should be slightly higher for higher $\pTjet$-triggered jets.
This would decrease the multiplicity of the parton shower at the initialization of the transport equation and reduce the amount of energy transfer to lower energy partons.

\subsection{Cone-size dependence of jet suppression and jet-shape modifications}
\label{sec:prediction:jetR-jetshape}
We also study the cone-size dependence of jet $R_{AA}$ and jet shape modifications that are closely related to the radial transport of jet energy.

\paragraph{Cone-size dependence of jet suppression}

\begin{figure}
    \centering
    \includegraphics[width=\textwidth]{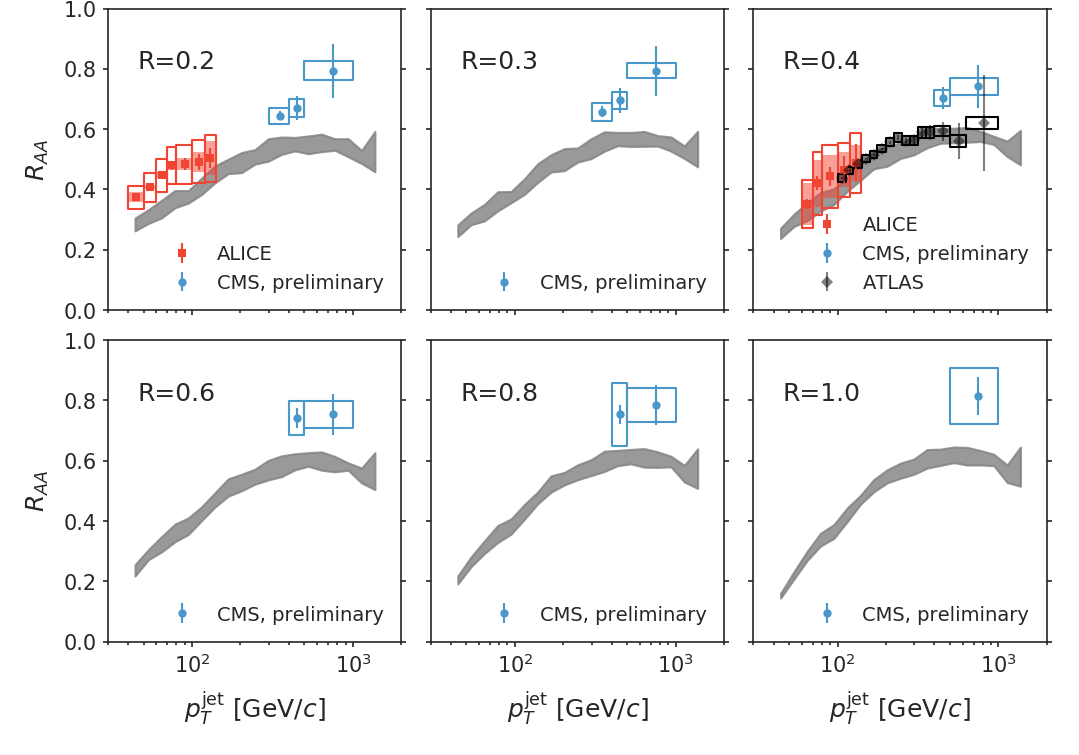}
    \caption{Predictions of cone-size dependence of inclusive jet nuclear modification factor in Pb-Pb collisions at $\sqrtsNN=5.02$ TeV, as compared to data from ALICE \cite{Acharya:2019jyg} and ATLAS \cite{Aaboud:2018twu} experiments, and preliminary data from the CMS experiment~\cite{CMS-PAS-HIN-18-014}.}
    \label{fig:jetRaa_R}
\end{figure}

\begin{figure}
    \centering
    \includegraphics[width=\textwidth]{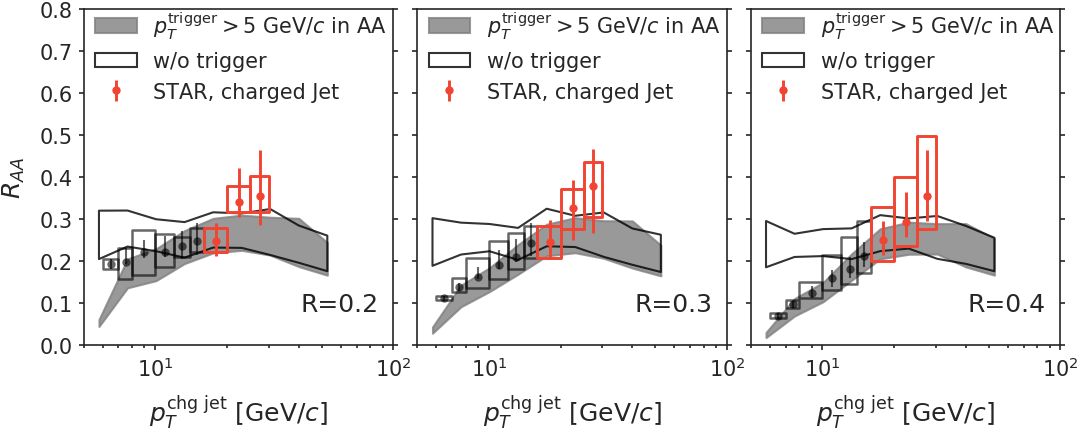}
    \caption{Predictions of cone-size dependence of inclusive jet nuclear modification factor in Au-Au collisions at $\sqrtsNN=200$ GeV, compared to charged-jet measurement from the STAR experiment~\cite{Adam:2020wen}. The experimental data colored in red (the three highest $\pTchgjet$ data points) are considered to be ``unbiased'', while the rest data points shown in back are subject to strong leading particle triggering bias.}
    \label{fig:jetRaa_R_RHIC}
\end{figure}

Jet reconstruction with a larger cone-size should recover more energy in the hard process.
The cone-size dependence of jet $R_{AA}$ comes from the competition of two effects: the increase of jet yield with $R$ in nuclear collisions and that in p-p collisions.
Recent measurements demonstrate that jet $R_{AA}$ has a weak $R$ dependence. 
At the RHIC energy, the STAR experiment observes that the charged jet $R_{AA}$ for $R=0.2, 0.3,$ and $0.4$ are the same within the experimental uncertainty \cite{Adam:2020wen}.
At the LHC energy, the ALICE measurement \cite{Acharya:2019jyg} shows similar $R_{AA}$ at $R=0.2$ and $R=0.4$.
the CMS measurement \cite{CMS-PAS-HIN-18-014} has extended jet measurement to $R=1.0$ for $\pTjet>400$ \GeVc, and the $R=1.0$ $R_{AA}$ shows less than 10\% increases compared to $R=0.2$ results.
This suggests energy lost at the core of the jet is transported mostly to very large angles.

In figures \ref{fig:jetRaa_R} and \ref{fig:jetRaa_R_RHIC}, the model calculations are compared to both the RHIC and the LHC data on jet $R_{AA}$ with different jet cone-sizes.
We notice that the CMS jet $R_{AA}$ at $R=0.4$ is about 0.15 higher than the ATLAS measurement in the similar $p_T$ and rapidity window, though the discrepancy is only at 1-2 $\sigma$ level.
Because we calibrate the model to ATLAS data at $R=0.4$, the predictions are systematically lower than the CMS measurements.
The model, however, explains the weak cone-size dependence at both the LHC and RHIC. 
A weak cone-size dependence is also found in the strongly-coupled hybrid approach to jet modification \cite{Pablos:2019ngg}.
Another weakly-coupled transport equation calculation predicts a stronger cone-size dependence \cite{He:2018xjv}.

The STAR experiment uses a leading particle trigger of $\pTtrigger>5$ GeV to define jets in Au-Au collisions to suppress ``fake jets'' from the background.
This trigger inevitably biases the reconstructed inclusive jet spectra.
Especially, the jet yield vanishes below the momentum of the trigger particle $\pTchgjet<\pTtrigger$.
Jet with higher transverse momentum is less affected by the trigger and the STAR experiment has estimated that the measured $R_{AA}$ with $\pTchgjet>16$ GeV, denoted by red symbols, are effectively free from the trigger bias.
For this reason, we only include the ``unbiased'' data points in the calibration in section \ref{app:bayes}.
Nevertheless, we can estimate the effect of trigger bias by comparing the inclusive jet $R_{AA}$ calculated without triggering (open solid box) to that with leading-particle triggering (gray bands).
With the triggering, the charged-jet $R_{AA}$ is almost flat as a function of $\pTchgjet$.
Our calculation with and without triggering agrees well above $\pTchgjet\sim 15$---$20$ GeV and diverges for lower momentum jet, which is in agreement with the estimation from the STAR experiment that $R_{AA}$ values above 16 GeV are effectively unbiased.
The calculation with trigger explains the dropping of the charged-jet $R_{AA}$ below $20$ GeV.
The calculation also shows that jets with smaller cone-size are less sensitive to the trigger bias because hard particles satisfying the momentum cut are more likely to be contained in jets with smaller cone-size at fixed $\pTchgjet$.

\paragraph{Jet shape}
Jet shape directly measures the radial distribution of energy around the jet.
Following the CMS measurement~\cite{Sirunyan:2018jqr}, we define the radial distribution of the transverse momentum of charged particles associated with $R=0.4$ jets 
\begin{eqnarray}
P(r) = \frac{1}{N_{\textrm{jets}}} \sum_{\textrm{jets}}\frac{\sum_{|r'-r|<\Delta r/2}  p_T}{\Delta r}.
\end{eqnarray}
The nested summation goes over charged particles for each jet.
Then, the jet shape is defined as a self-normalized transverse momentum distribution within $r<1$.
\begin{eqnarray}
\rho(r) = \frac{P(r)}{\int_0^1 P(r) dr}.
\end{eqnarray}
Note that there is a small but finite kinematic cut $\pTtrack>0.7$ \GeVc on charged particles in the CMS measurement.

\begin{figure}
    \centering
    \includegraphics[width=.5\textwidth]{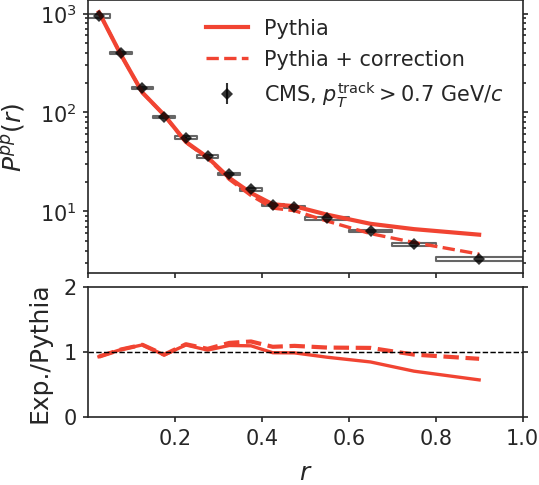}
    \caption{Pythia8 simulations of inclusive jet  shape in proton-proton collision (red-solid line), compared to data from the CMS experiment \cite{Sirunyan:2018jqr}. The red dashed line corresponds to a ``corrected'' simulation to better agree with the CMS data at large $r$. The ``corrected'' results has been subtracted by a 2.2 GeV per radian per rapidity background.}
    \label{fig:jetshape-baseline}
\end{figure}

In figure \ref{fig:jetshape-baseline}, the $p_T$ profile $P(r)$ in p-p collisions simulated by \pythia is compared to CMS data for $\pTjet>120$ \GeVc.
Inside the jet cone of $r<0.4$, the simulation (red solid line) describes the data well.
For $r>0.4$, \pythia simulations overshoot the data points, with a discrepancy about 50\% at $r=1.0$.
Such a deviation in the baseline can affect the interpretation
of the ratio between jet-shape in Pb-Pb and p-p.
For a fairer comparison in Pb-Pb, we try to use an {\it ad hoc} method to correct this discrepancy at large $r$.
Because the measured transverse momentum density is already small at large r: $P(r)/(2\pi r) \approx 0.55$ GeV per squared radian,
we attribute the discrepancy at large $r$ to a small mismatch in background estimation between experiment and simulation\footnote{For example, the comparison is sensitive to soft radiations and decay effects due to the finite kinematic cut on charged particles.}, $P_{\textrm{calc.}}(r)-P_{\textrm{exp.}}(r) = 2\pi r \epsilon_0$.
The fitted $\epsilon_0$ is around 0.37 GeV per squared radian. 
We then correct the simulated jet shape by subtracting this constant background. 
The corrected {\tt Pythia8} result is plotted as the red dashed line in figure \ref{fig:jetshape-baseline}, which agrees better with the measurement at a larger $r$.
Note that $\epsilon_0 \approx 0.37$ GeV hardly affects the jet energy determination as $\epsilon_0 \pi R^2 \approx$ 0.2 GeV is much less than the jet energy and energy loss.

\begin{figure}
    \centering
    \includegraphics[width=.55\textwidth]{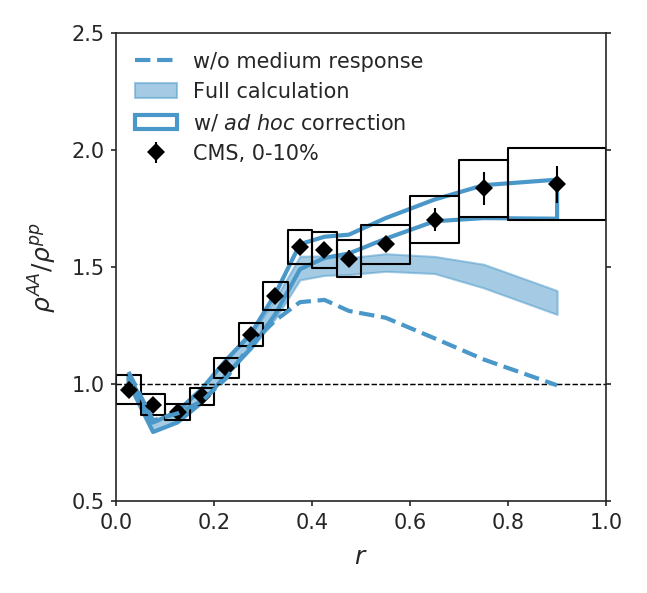}
    \caption{Prediction of the ratio between inclusive jet shape in Pb-Pb collisions and p-p collisions (blue band) compared to data from the CMS experiment \cite{Sirunyan:2018jqr}.
    The dashed line is the central prediction of jet shape without contributions from medium excitations.
     Blue open bands is the result corrected by the {\it ad hoc} formula in equation \ref{eq:correct_jetshape}, where a $\epsilon_0=0.35$ GeV per squared-radian energy density are subtracted from the transverse-momentum profile calculation for Pb-Pb and p-p.}
    \label{fig:jetshape}
\end{figure}

In figure \ref{fig:jetshape}, the ratio between jet shape $\rho(r)$ in Pb-Pb and that in p-p collisions is compared to CMS data.
The uncorrected prediction (blue band) describes well the dip of the ratio around $r=0.1$ and the fast-rising with increasing $r$.
At large radius, the prediction starts to deviate from the measurement where $\rho^{AA}/\rho^{pp} \approx 1.7$.
As we have assumed, this stems from a mismatch in the background definition.
We apply the {\it ad hoc} correction to the prediction
\begin{eqnarray}
\rho_c^{AA}(r)/\rho_c^{AA}(r) = \frac{N^{pp}}{N^{AA}}\frac{P^{AA}(r)-2\pi r\epsilon_0 }{P^{pp}(r)-2\pi r\epsilon_0},
\label{eq:correct_jetshape}
\end{eqnarray}
where the estimated mismatch in background $\epsilon_0$ is subtracted from both the p-p and Pb-Pb calculations, and $N^{AA}$ and $N^{pp}$ are the self-normalization factor of the corrected momentum profile $P(r)-2\pi r \epsilon_0$ for Pb-Pb and p-p collisions respectively.
The correction increases the calculated extended jet shape for $r>0.4$ to the measured level of ``energy excess'', shown as the open blue bands.
Finally, the dashed line shows the jet shape calculation where we exclude the contribution of medium excitations.
Comparing to the full model calculation, we conclude that medium excitation is negligible at small $r$.
At $r\sim 1$, its contribution is essential to explain the large-$r$ excess of jet shape modification.

\begin{figure}
    \centering
    \includegraphics[width=.5\textwidth]{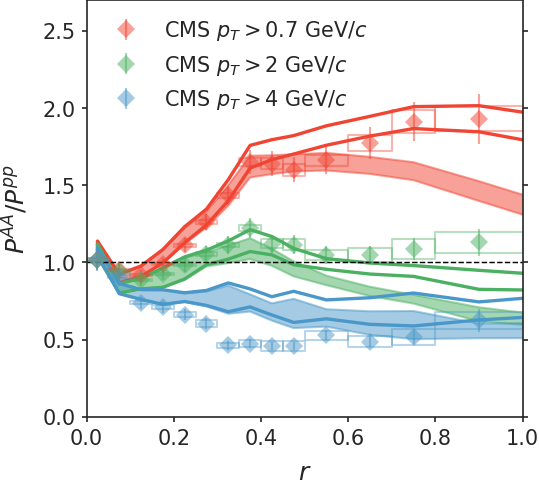}
    \caption{Prediction of the $p_T$-cut dependent jet shape compared to data from the CMS measurement \cite{Sirunyan:2018jqr}. The lines and symbols from top to bottom corresponds to the jet shape constructed using particles with $p_T>0.7, 2.0$, and 4.0 \GeVc respectively.
    Filled color bands are direct model predictions and open color bands are predictions corrected by the {\it ad hoc} procedure.
    }
    \label{fig:jetshape_pTmin}
\end{figure}

Modifications of jet fragmentation function and jet shape only contain information of the one-dimensional projection of the two-dimensional energy distributions $dE/(dp_Tdr)$ and $dE(d\zjet dr)$, and a direct comparison to the latter impose unprecedentedly detailed constraints to the modeling of jet-medium interaction.
For this purpose, the CMS experiment has measured the jet shape with different cuts on the transverse momentum of charged particles.

In figure \ref{fig:jetshape_pTmin}, we show the ratios of the transverse momentum profile $P(r)$ for charged particles with $p_T>0.7, 2.0$, and $4.0$ \GeVc between Pb-Pb and p-p collisions (filled bands) and compare to data.
The open bands are the ``corrected'' ratio as calculated in equation \ref{eq:correct_jetshape} where a constant transverse momentum background is subtracted so that the radial profile of transverse momentum density in proton-proton collisions agree with the data outside of the jet cone.
The residual background levels that we subtracted from the transverse-momentum distribution are $\epsilon_0 = 0.37,0.29$, and 0.12 GeV for $p_T$ cuts equals $0.7, 2.0$, and 4.0 \GeVc, respectively. 
After the correction, we see a better description of the extended jet shape at large $r$.
Outside the $R=0.4$, increasing the $p_T$ cut from $0.7$ \GeVc to $2$ \GeVc reduces the transverse-momentum outside the jet cone by about 50\%.
This can be understood from the simple test shown on the left of figure \ref{fig:PsiR_full} that the large-$r$ enhancement is dominated by soft $p_T \sim 2$ \GeVc particles and an increased $p_T$ cut effectively removes such contributions.
This effect is more prominent with $p_T$ cut increased to 4 \GeVc.

\section{Summary and outlook}
\label{sec:conclusion}
In this work, we have performed a transport-model based study of jet production and modifications in relativistic heavy-ion collisions.
Initial hard processes and vacuum-like parton showers are generated by {\tt Pythia8}, jet-medium interactions are treated in a previously developed linearized partonic transport model {\tt LIDO}, and the jet-induced medium excitation contains both hard parton recoils and a linearized hydrodynamic response.

Using data in single inclusive jet and single hadron suppression in central nuclear collisions at the LHC and RHIC, we systematically calibrated four model parameters: scale parameter $\mumin$ for the in-medium coupling, separation scale parameter $Q_0$ between vacuum-like and in-medium parton shower (separately tuned at the RHIC and the LHC), and a termination temperature $T_f$ for jet-medium interactions.
The calibrated model consistently describes inclusive jet and hadron suppression at high-$p_T$.
Parameter sets for central prediction and error estimation are reported.

\begin{figure}
    \centering
    \includegraphics[width=.8\textwidth]{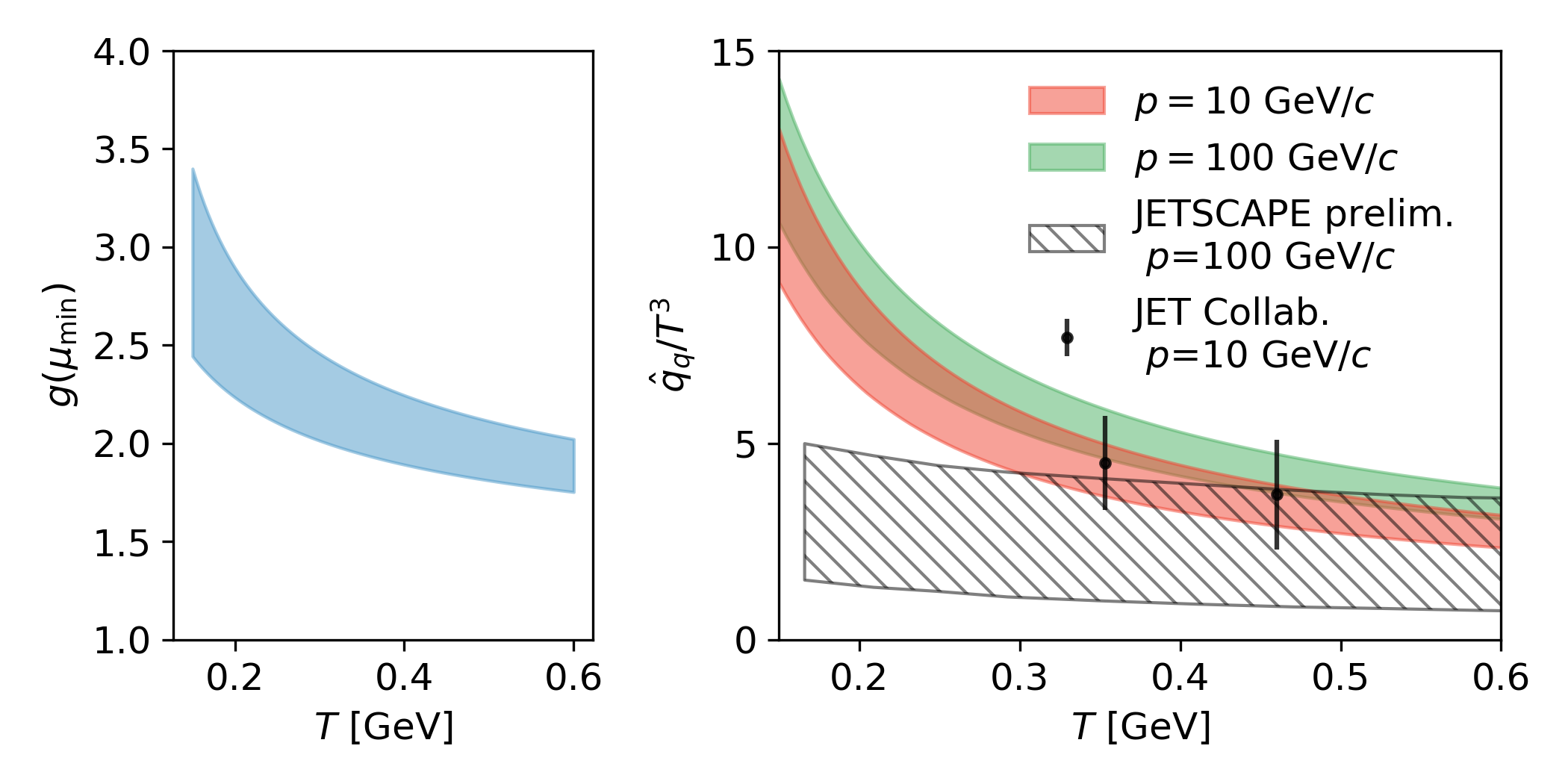}
    \label{fig:g_qhat}
    \caption{The resultant 95\% credible region of medium coupling strength $g(\mumin)$ as a function of temperature $T$ (left) and jet transport coefficient for light quark $\qhat_q$ scaled by $T^3$ (right). Red and green bands correspond to the value of $\qhat/T^3$ for a light quark with 10 \GeVc and 100 \GeVc momentum respectively. Black symbols with error bars are values estimated by the JET Collaboration \cite{PhysRevC.90.014909} and the hatched band is one of the preliminary results extracted by the JETSCAPE Collaboration \cite{Soltz:2019aea}
    }.
\end{figure}

The calibration also extracts of the effective medium-coupling strength $g_{\textrm{med}}=g(\mumin)$ (left) and the jet transport parameter $\qhat(p, T)$ (right) from both jet and hadron suppression data as shown in figure \ref{fig:g_qhat}.
The large calibrated values of the effective coupling at the medium scale $g_{\textrm{med}}\sim 2$--$3$ stands out as a potential problem because it challenges the use of weak-coupling assumption for jet-medium interaction.
We will revisit this problem in future studies with non-perturbative modeling of the medium constituents.
The extracted 95\% credible region of the scaled jet transport parameter $\qhat(p, T)/T^3$ of a light quark is consistent to earlier estimation by the JET Collaboration \cite{PhysRevC.90.014909}, but is larger than a recent extraction (preliminary) by the JETSCAPE Collaboration \cite{Soltz:2019aea}.
A possible explanation is that the JETSCAPE multi-stage approach includes medium effects in both the high-virtuality evolution and the transport equations of parton showers, but the present study accounts for all the jet-medium interactions in the transport equation.
As a result, the present study may have over-estimated the magnitude of the jet-medium coupling to compensate for the neglected medium effects in the evolution of high-virtuality partons.
In the future, we seek to include the {\tt LIDO} model as an alternative transport model in the JETSCAPE framework to benefit from the more advanced implementation of high-virtuality parton evolution in the medium using the MATTER event generator.

In the second half of this paper, we first analyzed the energy flow induced by jet-medium interactions to help understand the observed jet modifications.
We found that a medium-induced radiative process transfers energy from the large-$\zjet$ partons down to partons with momentum $p\sim \omega_{BH} \approx \pi T$.
Elastic energy loss and jet-induced medium excitation further redistribute the energy that has been accumulated around $\omega_{BH}$.

Using the calibrated model, we made predictions for modifications of jet fragmentation function, cone-size dependence of jet $R_{AA}$, and modification of jet shape.
The calibrated model produces a good agreement with the measured fragmentation function of inclusive jets at the LHC in central Pb-Pb collisions. 
The excess of low-$p_T$ fragmentation function and jet shape at  large-$r$ are found to be consequences of both medium-induced radiation and jet-induced medium excitation.

In conclusion, the transport-based model can simultaneously explain inclusive jet and hadron suppression at high-$p_T$.
The same model with calibrated parameters yields comparable jet-transport coefficient to earlier extractions and provides consistent descriptions of inclusive jet modifications including jet fragmentation function and jet shape.
These observations support the perturbative-based transport modeling of jet-medium interactions.

\acknowledgments
We thank Peter Jacobs, Robert L\'{i}\v{c}en\'{i}k, and Jana Biel\v{c}\'{i}kov\'{a} for helpful discussion on the STAR measurement of charged jet suppression. 
We thank Abhijit Majumder and Chun Shen for a thorough discussion and helpful feedback on this work.
This work was supported by DOE under Grant No. DE-AC02-05CH11231, by NSF under Grant No. ACI-1550228 within the JETSCAPE Collaboration, and by NSFC under Grant Nos. 11935007, 11221504, 11890714, and 11861131009, and by the UCB-CCNU Collaboration Grant. 
This work used resources of the National Energy Research Scientific Computing Center (NERSC), a U.S. Department of Energy Office of Science User Facility operated under Contract No. DE-AC02-05CH11231.

\appendix
\section{Medium evolution}
\label{app:medium}

We simulate the medium evolution using a hydrodynamic-based {\tt hic-eventgen} package that is described in detail in \cite{Bernhard:2018hnz}.
The package assumes boost-invariance and includes a parametric initial condition~\cite{Moreland:2014oya}, a free-streaming model for pre-equilibrium dynamics \cite{Broniowski:2008qk}, the 2+1D relativistic viscous hydrodynamics~\cite{Song:2007ux,SHEN201661}\footnote{
It also includes a particlization model that samples hadrons from the fluid and a hadronic afterburner. 
We only include jet-medium interaction in the QGP phase and neglects jet interaction with hadronic matter.
}.
For simplicity, we use event-averaged initial conditions to generate events for 0-10\% central nuclear collisions.
This is an acceptable approximation considering the observables that we consider, inclusive hadron and jet suppression, are most sensitive to long-wavelength structures of the medium.

We only extract the equilibrium information from the medium for the simulations of jet evolution.
This means that during the hydrodynamic stage, jet partons interact with a medium in local equilibrium, which is comprised of massless quarks and gluons, whose energy density $\epsilon$ and flow velocity $u^{\mu}$ match the viscous hydrodynamic simulation.
The temperature is defined using the lattice equation of state $T=T(\epsilon)$ \cite{Bazavov:2014pvz}.

\begin{figure}
    \centering
    \includegraphics[width=.9\textwidth]{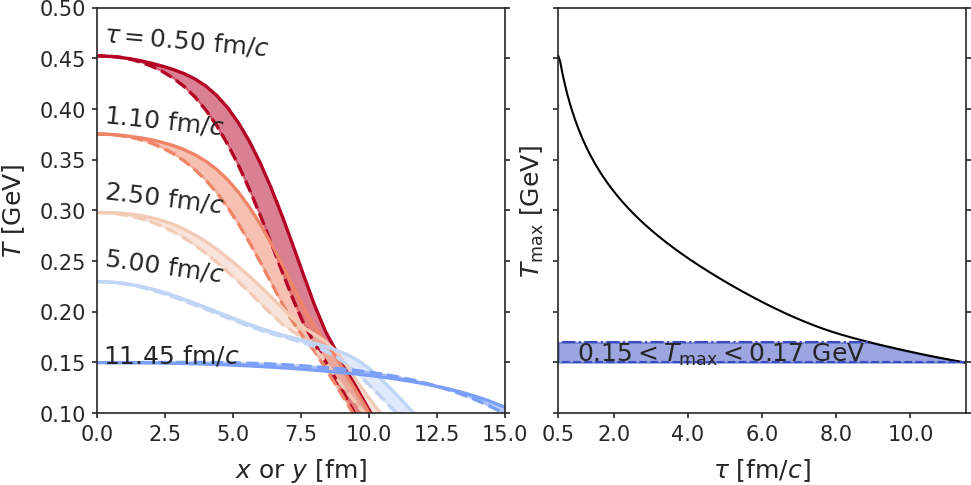}
    \caption{Left: snapshots of the temperature profiles at different time of the QGP fireball from the hydrodynamic simulation for 0-10\% central Pb-Pb collisions at $\sqrtsNN=5.02$ TeV. 
    Dashed (solid) lines plot temperature variations from the center along the short (long) axis of the fireball. 
    Right: the maximum temperature of the medium is plotted as a function of proper time. The band shows the range of variation of termination temperature $T_f$ of jet-medium interaction in the calibration.}
    \label{fig:hydro}
\end{figure}

In figure \ref{fig:hydro}, we plot the temperature profiles (left) and the evolution of the maximum temperature with proper time (right).
From the right plot, we see that a 0.02 GeV decrease in $T_f$ leads a longer duration of jet-medium interaction: $\Delta \tau\approx 2.5$ fm/$c$.

\section{Comparisons of single-gluon emission rates}
\label{app:validate_lido}

\begin{figure}
\centering
\includegraphics[width=.48\textwidth]{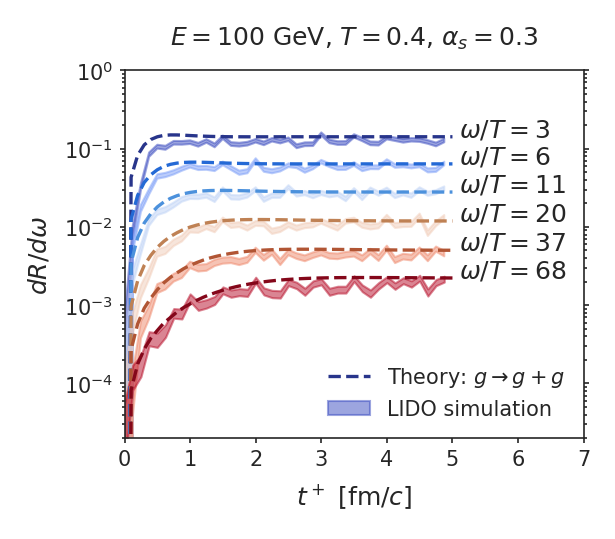}
\includegraphics[width=.48\textwidth]{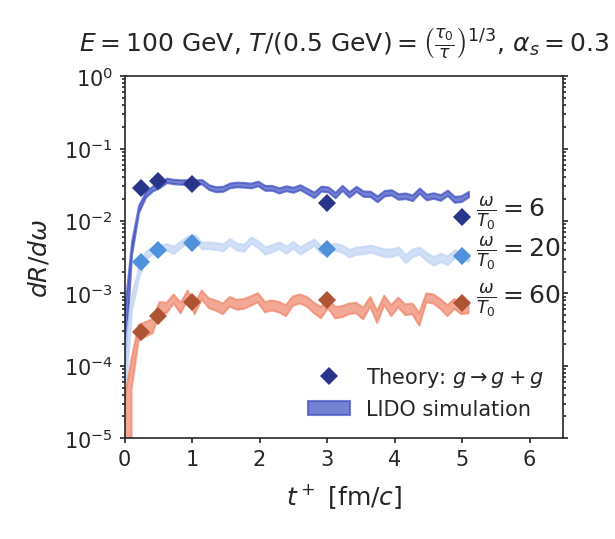}
\caption{Comparisons of the full theory calculations (dashed lines and symbols) of the single splitting rate of the process $g\rightarrow g+g$ to the {\tt LIDO} simulations (bands). 
Left: differential rate $dR/d\omega$ in a static medium as a function of patron evolution time $t^+$, plotted at different energy of the daughter gluon: $\omega /T=3,6,11,20,37,68$.
Right: differential rate $dR/d\omega$ in ax expanding medium as a function of patron evolution time $t^+$, plotted at different energy of the daughter gluon: $\omega/T=6,20,60$. The medium expands according to Bjorken flow where $T^3\propto \tau^{-1}$.}
\label{fig:theo_to_simu}
\end{figure}

We shall demonstrate in this appendix that the {\tt LIDO} model is able to quantitatively reproduce the single parton emission rate as calculated in the full theory \cite{Zakharov:1996fv,Zakharov:1997uu,CaronHuot:2010bp}, where multiple collisions to the single parton emissions are resummed to all orders with full leading-order weakly-coupled jet-medium interaction.
In \cite{Ke:2018jem}, we had compared our simulation to the published numerical results in reference \cite{CaronHuot:2010bp} for a quark at relatively low energy $E=16$ GeV.
In this work, we use a new implementation \cite{Ke:2020coj} of the approach introduced in \cite{CaronHuot:2010bp} to solve the single parton radiation rate from the full theory numerically.
It allows us to compare our simulation to the theory calculation at higher parton energy that is relevant for jets at the LHC energies and also extend to the case of an expanding medium. 
In this appendix, we choose to demonstrate this comparison using the $g\rightarrow g+g$ splitting channel with the initial gluon energy $E=100$ GeV.
Both the simulation and the numerical solution of the theory use a fixed coupling constant $\alpha_s=0.3$.

On the left panel of figure \ref{fig:theo_to_simu}, we compare the differential rate (bands) $dR/d\omega$ of gluon splitting $g\rightarrow g+g$ as a function of time at which the radiation takes place in a finite and static medium.
$omega$ is the energy of one of the radiated gluons.
Bands and dashed lines correspond to {\tt LIDO} simulations and theoretical calculations, respectively.
Different colors label the differential rates at different gluon energies.
We can see that simulations quantitatively agree with the theoretical calculations at late times.
The model also captures the radiation rate in the transition region at early times, which is critical for the model to reproduce the non-linear path-length dependent radiative energy loss of leading partons.
This agreement is better for energetic daughter parton with $\omega/T\gg 1$.
At lower energy $\omega \gg 3T$, one observes notable deviations at early times.

On the right panel of figure \ref{fig:theo_to_simu}, we make the comparison between theory (symbols) and simulations (bands) in an expanding medium. 
The temperature of the medium drops according to that of the Bjorken flow \cite{Bjorken:1982qr}, where $T^3 \propto \tau^{-1}$ with an initial temperature $T_0=0.5$ GeV at the initial time of $\tau_0=0.5$ fm/$c$.
We observe that simulations capture the time dependence of the differential radiation rate very well for high energy daughter gluons. 
However, the current model produces more low-energy daughter gluons at late times than what is expected from the theory.

One concludes that in both cases, the agreement between theory and simulation is satisfactory for those emitting gluons with thermal energies $\omega \sim 3T$.
We do not worry too much about this discrepancy at this moment for two reasons.
First, for a moderate coupling $\alpha_s = 0.3$, the screening mass is of the same order as the kinetic energy of a gluon with thermal energy. 
The collinearity between the final-state and initial-state partons as the theoretical formula assumed breaks down.
Therefore, the theory calculation already approaches its limits of applicability.
Second, emissions of gluons with thermal momenta do not strongly affect the radiative energy loss of the leading parton, and their impact on soft partons are overwhelmed by those caused by elastic collisions.
Therefore, despite the current discrepancy between our model and the full theory around $\omega\sim 3T$, we do not expect it will strongly affect the prediction on physical observables.

\section{Details of the Bayesian model parameter calibration}
\label{app:bayes}

\subsection{Procedures of Bayesian parameter calibration}
We refer the readers to the reference \cite{Bernhard:2018hnz} for a detailed review of the application of the Bayesian method to model calibration in the context of relativistic heavy-ion physics.
Specific to our application, we perform the Bayesian model calibration in the following steps.
\paragraph{Model computation at the design parameter points} 
Fifty parameter sets are sampled using the Latin-hypercube sampling procedure from the four-dimensional parameter space 
\begin{eqnarray}
\mathbf{p}=\left[ \ln\frac{\mumin}{\pi T}, \ln\frac{\QLHC}{1~\textrm{GeV}},
\ln\frac{\QRHIC}{1~\textrm{GeV}},
T_f
\right]
\end{eqnarray}
with the range introduced in section \ref{sec:bayes:parameter}.
The Latin-hypercube sampling algorithm is implemented in the {\tt R} package \cite{R,R-LHS}.
Each of these fifty parameter points, $\mathbf{p}_i$, predicts a set of inclusive jet and hadron $R_{AA}$ through the full model simulation.
The list of $R_{AA}$ for jet and hadron at different colliding energies is concatenated into a single training observable vector $\mathbf{y}_i$ for the parameter point labeled by $i$.

\paragraph{Dimensional reduction of the training data}
We use principal component analysis, as implemented in the {\tt scikit-learn} \cite{scikit-learn} package, to compress the information of the observable vectors $\mathbf{y}_i$ into vectors of lower dimensional $\mathbf{z}_i$ where $\mathrm{dim} \mathbf{z} \ll \mathrm{dim} \mathbf{y}$. 
First, the reduction of the dimension of the observable is possible because many outputs vary in a highly correlated manner when parameters are varied.
For example, in our study, an increase in coupling strength decreases all high-$p_T$ $R_{AA}$.
Second, the outputs are dominated by linear correlations.
For example, the amount of change in jet $R_{AA}$ approximately scales linearly with the change in inclusive hadron $R_{AA}$, when we vary the model parameters. 
In this case, the change in the full observable vectors can be approximately decomposed into linear combinations of just a few bases vectors $\mathbf{b}_j$ labeled by $j$
\begin{eqnarray}
\mathbf{y}_i = \sum_{j=1}^{N_{\textrm{pc}}}z_{ij} \mathbf{b}_j + (\textrm{Truncation errors}).
\end{eqnarray}
Now, each row of $z_{ij}$ becomes the transformed observable vector for parameter set ``$i$'', and different columns of $z$ are guaranteed to be linearly independent. 
When the dimension of $z$--$N_{\textrm{pc}}$--is chosen to be smaller than $\dim \mathbf{y}$, the linear combination does not fully agree with the original vector and introduces finite truncation errors.
For a well-behaved model with a few parameters, a small number of $N_{\textrm{pc}}$ is enough to reproduce the original vector with small truncation errors.
We choose $N_{\textrm{pc}}=5$ in this work, which accounts for 97.5\% of the variation of the original data.
Variances in the truncated data are propagated to the interpolation uncertainty as white noises.

\paragraph{Emulate model prediction by interpolating training data.}
Next, we define an interpolation $\mathbf{z}=f(\mathbf{p})$ to obtain model prediction $\mathbf{z}$ at a general point $\mathbf{p}$ in the parameter space. 
The interpolation is realized by training the Gaussian processes on the fifty training (design) parameter sets and the corresponding predicted observable vectors.
A Gaussian-process emulator is a procedure to sample a random $n$-dimensional scalar function whose outputs at different inputs follow a multivariate Gaussian distribution.
The training procedure adapts the hyper-parameters in the Gaussian-process emulator so that the distribution of the generated random functions is conditioned to the training output $\mathbf{z}_i$ at training inputs $\mathbf{p}_i$.
At a novel input point $\bfx_i$, the ensemble of random function values defines the central prediction and uncertainty of the interpolation.
Finally, we transform the predicted $\mathbf{z}$ back to predictions of physical observables $\mathbf{y}$.
The prediction uncertainty that is represented by a diagonal covariance matrix in the principal component space is also transformed back to the physical space $\Sigma_{GP}$.
All these functionalities of building and training Gaussian process emulators are provided by the {\tt scikit-learn} package \cite{scikit-learn}.

\paragraph{Model discrepancy}
It is hardly true that a model, even with optimal parameters, can accurately describe every detail of an observable. 
This is the so-called model discrepancy.
Without accounting for the model discrepancy in the uncertainty estimation, a calibration potentially leads to over-confident constraints on parameters.
In reality, for a complex model as in this work, it is impossible to quantify the discrepancy between the model and the true underlying theory.
We adopt a strategy of including model discrepancy used in \cite{Bernhard:2018hnz}.
In this method, a relative uncertainty $\sigma_{\textrm{model}}$ is assigned to the prediction uncertainty of each principal component in addition to the uncertainty of the Gaussian-process emulators.
Because we do not know the exact value of $\sigma_{\textrm{model}}$, we treat it as another parameter to be marginalized in the Bayesian parameter calibration.
Moreover, we give $\sigma_{\textrm{model}}$ an informative prior,
\begin{eqnarray}
P_0(\sigma_{\textrm{model}}) \propto \sigma_{\textrm{model}}^2 e^{-\sigma_{\textrm{model}}/0.05}, 0<\sigma_{\textrm{model}}<0.5
\end{eqnarray}
This results in a 15\% average model discrepancy prior to the model-to-data comparison.

\paragraph{Model-to-data comparison from Bayes' theorem}
Assisted by a fast model emulator, we can infer model calculations with uncertainty at any point in the considered range of parameters.
Next, we apply Bayes' theorem to combine the model and experimental data to define the posterior probability distribution of parameters $P(\mathbf{p})$.

\begin{eqnarray}
P(\mathbf{p}) = \exp\{-\frac{1}{2}\delta\mathbf{y}_{\mathbf{p}}\Sigma^{-1}\mathbf{y}_{\mathbf{p}}^T\}P_0(\mathbf{p})
\label{eq:posterior-1}
\end{eqnarray}
where $\mathbf{\delta y}_{\mathbf{p}} = \mathbf{y}(\mathbf{p})-\mathbf{y}_{\exp}$. 
$\Sigma$ is the uncertainty matrix that combines interpolation uncertainty and experimental uncertainty.
\begin{eqnarray}
\Sigma = \Sigma_{GP}+\Sigma_{\textrm{model}} + \mathrm{diag}\{\left(\delta \mathbf{y}_{\exp}^{\textrm{stat}}+\delta \mathbf{y}_{\exp}^{\textrm{sys}} \right)^2\}
  + \sum_{\exp} \oplus (\delta \mathbf{y}_{\exp}^{\textrm{norm}} \otimes \delta \mathbf{y}_{\exp}^{\textrm{norm}}),
  \label{eq:posterior-2}
\end{eqnarray}
where $\Sigma_{GP}$ is the covariance matrix of the interpolation procedure. $\delta \mathbf{y}_{\exp}^{\textrm{stat}}+\delta \mathbf{y}_{\exp}^{\textrm{sys}}$ is the experimental statistical uncertainty plus systematic uncertainty, which only contributes to the diagonal term of $\Sigma$.
In reality, systematic uncertainties usually have a certain degree of correlation.
For simplicity, they are treated as uncorrelated in this analysis except for the normalization uncertainty.
The normalization uncertainty $\delta \mathbf{y}_{\exp}^{\textrm{norm}}$ comes from binary collision $T_{AA}$ uncertainty and luminosity uncertainty in $R_{AA}$ measurement; therefore, they are fully-correlated uncertainty over different transverse momentum bins in one measurement.
The $\otimes$ denotes the outer product of the uncertainty array for one of the experiments, and the $\oplus$ denotes the direct sum of normalization uncertainty covariance matrices among different experiments.
The normalization uncertainty are included as percentage of the central reported experimental values $\delta \mathbf{y}_{\exp}^{\textrm{norm}} = \textrm{norm} \times \delta \mathbf{y}_{\exp}$.
The normalization factor for different experiments in 0-10\% Pb-Pb or 0-10\% Au-Au collisions are listed in table \ref{tab:exp-norm}.
To obtain the maginalized posterior distribution that has been discussed in section \ref{sec:bayes:posterior}, we use Markov chain Monte Carlo (MCMC), as implemented in the {\tt emcee} package \cite{emcee}, to pull random samples from the posterior distribution defined in equations \ref{eq:posterior-1} and \ref{eq:posterior-2} and project onto the one- or two-parameter subspace.

\begin{table}[ht]
    \centering
    \begin{tabular}{cccccc}
    \hline
        ATLAS jet &  ALICE jet & CMS chg. part. & CMS $D^0$ & STAR chg. jet & PHENIX $\pi^0$\\
        \hline
       0.0085 & 0.028 & 0.025 & 0.025 & 0.07 & 0.12\\
        \hline
    \end{tabular}
    \caption{Normalization uncertainty level of different measurements in 0-10\% Pb-Pb collision or 0-10\% Au-Au collisions.}
    \label{tab:exp-norm}
\end{table}

\begin{figure}
    \centering
    \includegraphics[width=.8\textwidth]{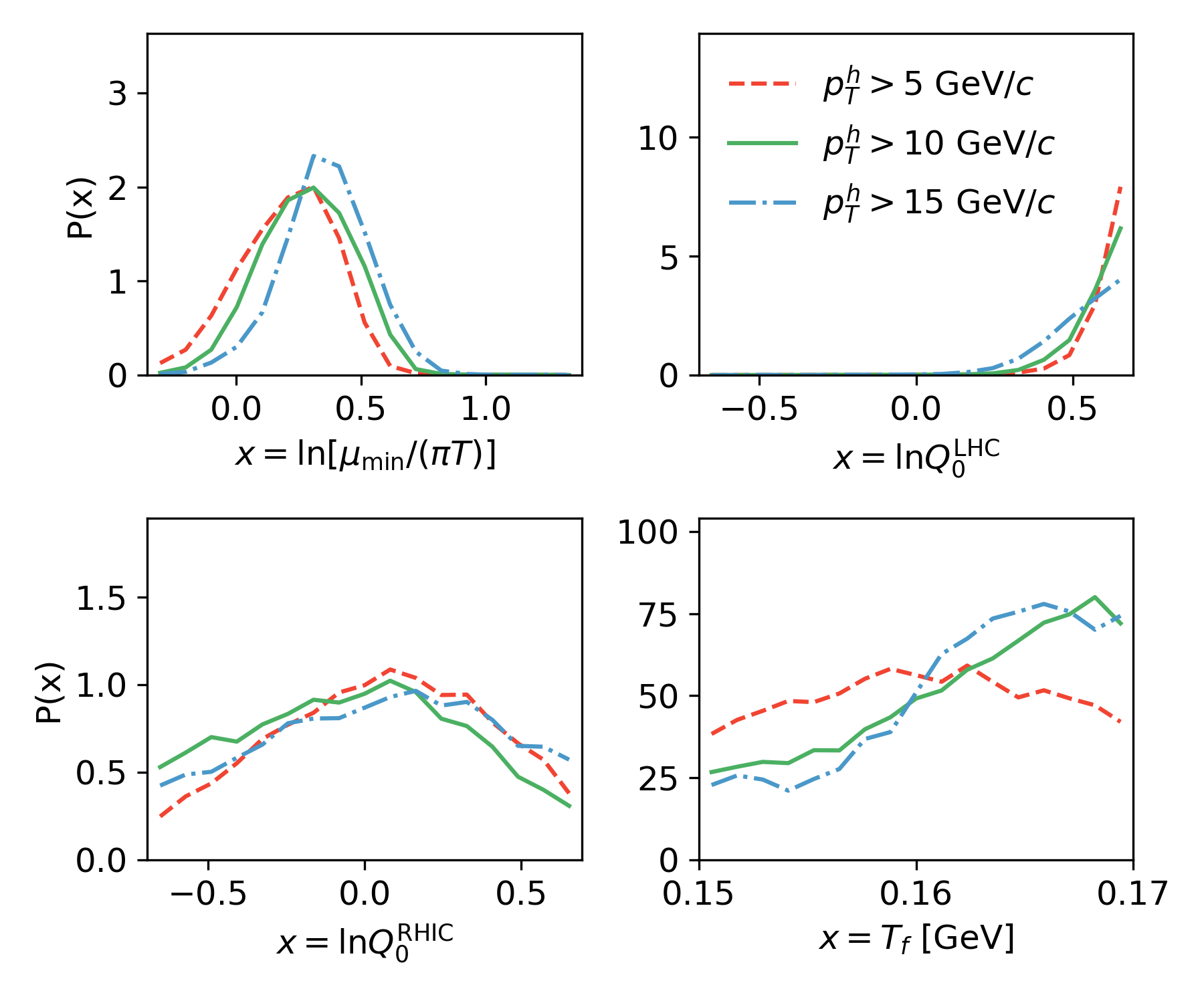}
    \caption{Test the sensitivity of posterior to the $p_T$ cut-off on hadron $R_{AA}$ data. We only show single-parameter posterior. Different color and line-shape combinations correspond to different values of cut-off $\pThadron>5$, 10, and 15 \GeVc respectively.}
    \label{fig:checkpT}
\end{figure}

\subsection{Sensitivity to the \texorpdfstring{$\pThadron$}~ cut on hadron suppression data}
\label{app:bayes:pTcut}
In section \ref{sec:bayes:posterior}, we only include single hadron suppression data above $p_T=10$ \GeVc.
The value of this cut-off momentum can be subjective, so we test the impact of a different cut-off $p_T$ on the posterior.
In figure \ref{fig:checkpT}, we compare three single-parameter posterior distributions for different choices of hadron data $p_T$ cut -- $\pThadron>5$ \GeVc (red dashed lines), $\pThadron>10$ \GeVc (green solid lines), and $\pThadron>15$ \GeVc (blue dash-dotted lines).
Including more low-$\pThadron$ data slightly decreases the preferred in-medium coupling scale, which increases the coupling strength.
It also drives the preferred termination temperature for jet-medium interaction toward lower values, increasing the effect of hot-medium on jet partons.
Both of these changes try to enhance the medium modifications on a single particle.
The matching scale parameter $Q_0$ at LHC energy favors higher values when we include more low-$\pThadron$ data.
A higher $Q_0$ effectively reduces vacuum showers, which compensates for the increased suppression on high-$p_T$ particles.
The posterior of $Q_0$ parameters at RHIC is consistent when changing $\pThadron$ due to large calibration uncertainty.

The posterior's dependence on $\pThadron$ cut-off on the calibration data set suggests that our model is not entirely consistent with the $p_T$ dependence of inclusive hadron data. As a result, lowering $\pThadron$ cut-off, the calibration has to balance the losses and gains with the low-$\pThadron$ data.
This is particular prominent for the $T_f$ parameter when we switch $\pThadron$ cut-off from 10 \GeVc to 5 \GeVc.
Despite this dependence on $\pThadron$, the three posterior distributions still largely overlap with each other.
In particular, the 95\% credible interval are consistent as listed in table \ref{tab:sets_pT}.
\begin{table}[ht]
    \centering
    \renewcommand{\arraystretch}{1.5}
    \begin{tabular}{ccccc}
    \hline
      $\pThadron$ cut-off   &
      $\ln\frac{\mumin}{\pi T}$ & $\ln\frac{\QLHC}{1~\textrm{GeV}}$ &
      $\ln\frac{\QRHIC}{1~\textrm{GeV}}$ &
      $T_f$ [GeV]\\ 
    \hline
      5 \GeVc  & 
      $0.222_{-0.421}^{+0.306}$ & $0.639_{-0.254}^{+0.052}$ & 
      $0.061_{-0.659}^{+0.564}$ &
      $0.160_{-0.010}^{+0.009}$
      \\
      10 \GeVc &
      $0.286_{-0.371}^{+0.331}$ &
      $0.612_{-0.275}^{+0.078}$ &
      $-0.023_{-0.620}^{+0.635}$ &
      $0.163_{-0.012}^{+0.007}$
      \\
      15 \GeVc &
      $0.358_{-0.381}^{+0.329}$ &
      $0.559_{-0.336}^{+0.127}$ &
      $0.060_{-0.691}^{+0.587}$ &
      $0.163_{-0.012}^{+0.006}$\\
    \hline     
    \end{tabular}
    \caption{The medians and 95\% credible limits of model parameters obtained for different choice of $\pThadron$ cut-off on experimental data.}
    \label{tab:sets_pT}
\end{table}

\bibliographystyle{JHEP}
\bibliography{ref}

\providecommand{\href}[2]{#2}\begingroup\raggedright\begin{thebibliography}{100}

\bibitem{PhysRevD.33.717}
D.~A. Appel, {\it Jets as a probe of quark-gluon plasmas},  {\em Phys. Rev. D}
  {\bf 33} (Feb, 1986) 717--722.

\bibitem{GYULASSY1990432}
M.~Gyulassy and M.~Plümer, {\it Jet quenching in dense matter},  {\em Physics
  Letters B} {\bf 243} (1990), no.~4 432 -- 438.

\bibitem{PhysRevLett.68.1480}
X.-N. Wang and M.~Gyulassy, {\it Gluon shadowing and jet quenching in a+a
  collisions at $\sqrt{s}=200a$ gev},  {\em Phys. Rev. Lett.} {\bf 68} (Mar,
  1992) 1480--1483.

\bibitem{Qin:2015srf}
G.-Y. Qin and X.-N. Wang, {\it Jet quenching in high-energy heavy-ion
  collisions},  {\em International Journal of Modern Physics E} {\bf 24}
  (2015), no.~11 1530014,
  [\href{http://arxiv.org/abs/https://doi.org/10.1142/S0218301315300143}{{\tt
  https://doi.org/10.1142/S0218301315300143}}].

\bibitem{Bass:2008rv}
S.~A. Bass, C.~Gale, A.~Majumder, C.~Nonaka, G.-Y. Qin, T.~Renk, and
  J.~Ruppert, {\it {Systematic Comparison of Jet Energy-Loss Schemes in a
  realistic hydrodynamic medium}},  {\em Phys. Rev. C} {\bf 79} (2009) 024901,
  [\href{http://arxiv.org/abs/0808.0908}{{\tt arXiv:0808.0908}}].

\bibitem{PhysRevC.90.014909}
{\bf JET Collaboration} Collaboration, K.~M. Burke, A.~Buzzatti, N.~Chang,
  C.~Gale, M.~Gyulassy, U.~Heinz, S.~Jeon, A.~Majumder, B.~M\"uller, G.-Y. Qin,
  B.~Schenke, C.~Shen, X.-N. Wang, J.~Xu, C.~Young, and H.~Zhang, {\it
  Extracting the jet transport coefficient from jet quenching in high-energy
  heavy-ion collisions},  {\em Phys. Rev. C} {\bf 90} (Jul, 2014) 014909.

\bibitem{Andres:2016iys}
C.~Andr\'es, N.~Armesto, M.~Luzum, C.~A. Salgado, and P.~Zurita, {\it {Energy
  versus centrality dependence of the jet quenching parameter $\hat{q}$ at RHIC
  and LHC: a new puzzle?}},  {\em Eur. Phys. J. C} {\bf 76} (2016), no.~9 475,
  [\href{http://arxiv.org/abs/1606.04837}{{\tt arXiv:1606.04837}}].

\bibitem{Xu:2017obm}
Y.~Xu, J.~E. Bernhard, S.~A. Bass, M.~Nahrgang, and S.~Cao, {\it {Data-driven
  analysis for the temperature and momentum dependence of the heavy-quark
  diffusion coefficient in relativistic heavy-ion collisions}},  {\em Phys.
  Rev. C} {\bf 97} (2018), no.~1 014907,
  [\href{http://arxiv.org/abs/1710.00807}{{\tt arXiv:1710.00807}}].

\bibitem{Xie:2019oxg}
M.~Xie, S.-Y. Wei, G.-Y. Qin, and H.-Z. Zhang, {\it {Extracting jet transport
  coefficient via single hadron and dihadron productions in high-energy
  heavy-ion collisions}},  {\em Eur. Phys. J. C} {\bf 79} (2019), no.~7 589,
  [\href{http://arxiv.org/abs/1901.04155}{{\tt arXiv:1901.04155}}].

\bibitem{Connors:2017ptx}
M.~Connors, C.~Nattrass, R.~Reed, and S.~Salur, {\it {Jet measurements in heavy
  ion physics}},  {\em Rev. Mod. Phys.} {\bf 90} (2018) 025005,
  [\href{http://arxiv.org/abs/1705.01974}{{\tt arXiv:1705.01974}}].

\bibitem{Chien:2015hda}
Y.-T. Chien and I.~Vitev, {\it {Towards the understanding of jet shapes and
  cross sections in heavy ion collisions using soft-collinear effective
  theory}},  {\em JHEP} {\bf 05} (2016) 023,
  [\href{http://arxiv.org/abs/1509.07257}{{\tt arXiv:1509.07257}}].

\bibitem{Casalderrey-Solana:2015vaa}
J.~Casalderrey-Solana, D.~C. Gulhan, J.~G. Milhano, D.~Pablos, and
  K.~Rajagopal, {\it {Predictions for Boson-Jet Observables and Fragmentation
  Function Ratios from a Hybrid Strong/Weak Coupling Model for Jet Quenching}},
   {\em JHEP} {\bf 03} (2016) 053, [\href{http://arxiv.org/abs/1508.00815}{{\tt
  arXiv:1508.00815}}].

\bibitem{Kang:2016ehg}
Z.-B. Kang, F.~Ringer, and I.~Vitev, {\it {Jet substructure using
  semi-inclusive jet functions in SCET}},  {\em JHEP} {\bf 11} (2016) 155,
  [\href{http://arxiv.org/abs/1606.07063}{{\tt arXiv:1606.07063}}].

\bibitem{Chang:2016gjp}
N.-B. Chang and G.-Y. Qin, {\it {Full jet evolution in quark-gluon plasma and
  nuclear modification of jet production and jet shape in Pb+Pb collisions at
  2.76ATeV at the CERN Large Hadron Collider}},  {\em Phys. Rev. C} {\bf 94}
  (2016), no.~2 024902, [\href{http://arxiv.org/abs/1603.01920}{{\tt
  arXiv:1603.01920}}].

\bibitem{Tachibana:2018yae}
{\bf JETSCAPE} Collaboration, Y.~Tachibana et~al., {\it {Jet substructure
  modifications in a QGP from multi-scale description of jet evolution with
  JETSCAPE}},  {\em PoS} {\bf HardProbes2018} (2018) 099,
  [\href{http://arxiv.org/abs/1812.06366}{{\tt arXiv:1812.06366}}].

\bibitem{Casalderrey-Solana:2018wrw}
J.~Casalderrey-Solana, Z.~Hulcher, G.~Milhano, D.~Pablos, and K.~Rajagopal,
  {\it {Simultaneous description of hadron and jet suppression in heavy-ion
  collisions}},  {\em Phys. Rev. C} {\bf 99} (2019), no.~5 051901,
  [\href{http://arxiv.org/abs/1808.07386}{{\tt arXiv:1808.07386}}].

\bibitem{Pablos:2019ngg}
D.~Pablos, {\it {Jet Suppression From a Small to Intermediate to Large
  Radius}},  {\em Phys. Rev. Lett.} {\bf 124} (2020), no.~5 052301,
  [\href{http://arxiv.org/abs/1907.12301}{{\tt arXiv:1907.12301}}].

\bibitem{Caucal:2020xad}
P.~Caucal, E.~Iancu, A.~Mueller, and G.~Soyez, {\it {Nuclear modification
  factors for jet fragmentation}},  \href{http://arxiv.org/abs/2005.05852}{{\tt
  arXiv:2005.05852}}.

\bibitem{Chen:2020tbl}
W.~Chen, S.~Cao, T.~Luo, L.-G. Pang, and X.-N. Wang, {\it {Medium modification
  of $\gamma$-jet fragmentation functions in Pb+Pb collisions at LHC}},  {\em
  Phys. Lett. B} {\bf 810} (2020) 135783,
  [\href{http://arxiv.org/abs/2005.09678}{{\tt arXiv:2005.09678}}].

\bibitem{Baier:1994bd}
R.~Baier, Y.~L. Dokshitzer, S.~Peigne, and D.~Schiff, {\it {Induced gluon
  radiation in a QCD medium}},  {\em Phys. Lett. B} {\bf 345} (1995) 277--286,
  [\href{http://arxiv.org/abs/hep-ph/9411409}{{\tt hep-ph/9411409}}].

\bibitem{BAIER1997265}
R.~Baier, Y.~Dokshitzer, A.~Mueller, S.~Peigné, and D.~Schiff, {\it Radiative
  energy loss and $p_{\perp}$-broadening of high energy partons in nuclei},
  {\em Nuclear Physics B} {\bf 484} (1997), no.~1 265 -- 282.

\bibitem{Zakharov:1996fv}
B.~Zakharov, {\it {Fully quantum treatment of the Landau-Pomeranchuk-Migdal
  effect in QED and QCD}},  {\em JETP Lett.} {\bf 63} (1996) 952--957,
  [\href{http://arxiv.org/abs/hep-ph/9607440}{{\tt hep-ph/9607440}}].

\bibitem{Zakharov:1997uu}
B.~Zakharov, {\it {Radiative energy loss of high-energy quarks in finite size
  nuclear matter and quark - gluon plasma}},  {\em JETP Lett.} {\bf 65} (1997)
  615--620, [\href{http://arxiv.org/abs/hep-ph/9704255}{{\tt hep-ph/9704255}}].

\bibitem{Baier:1998kq}
R.~Baier, Y.~L. Dokshitzer, A.~H. Mueller, and D.~Schiff, {\it {Medium induced
  radiative energy loss: Equivalence between the BDMPS and Zakharov
  formalisms}},  {\em Nucl. Phys. B} {\bf 531} (1998) 403--425,
  [\href{http://arxiv.org/abs/hep-ph/9804212}{{\tt hep-ph/9804212}}].

\bibitem{Wiedemann:2000za}
U.~A. Wiedemann, {\it {Gluon radiation off hard quarks in a nuclear
  environment: Opacity expansion}},  {\em Nucl. Phys. B} {\bf 588} (2000)
  303--344, [\href{http://arxiv.org/abs/hep-ph/0005129}{{\tt hep-ph/0005129}}].

\bibitem{Gyulassy:2000er}
M.~Gyulassy, P.~Levai, and I.~Vitev, {\it {Reaction operator approach to
  nonAbelian energy loss}},  {\em Nucl. Phys. B} {\bf 594} (2001) 371--419,
  [\href{http://arxiv.org/abs/nucl-th/0006010}{{\tt nucl-th/0006010}}].

\bibitem{Arnold:2002ja}
P.~B. Arnold, G.~D. Moore, and L.~G. Yaffe, {\it {Photon and gluon emission in
  relativistic plasmas}},  {\em JHEP} {\bf 06} (2002) 030,
  [\href{http://arxiv.org/abs/hep-ph/0204343}{{\tt hep-ph/0204343}}].

\bibitem{THOMA1991128}
M.~H. Thoma, {\it Collisional energy loss of high energy jets in the
  quark-gluon plasma},  {\em Physics Letters B} {\bf 273} (1991), no.~1 128 --
  132.

\bibitem{PhysRevD.44.R2625}
E.~Braaten and M.~H. Thoma, {\it Energy loss of a heavy quark in the
  quark-gluon plasma},  {\em Phys. Rev. D} {\bf 44} (Nov, 1991) R2625--R2630.

\bibitem{Mustafa:2003vh}
M.~G. Mustafa and M.~H. Thoma, {\it {Quenching of hadron spectra due to the
  collisional energy loss of partons in the quark gluon plasma}},  {\em Acta
  Phys. Hung. A} {\bf 22} (2005) 93--102,
  [\href{http://arxiv.org/abs/hep-ph/0311168}{{\tt hep-ph/0311168}}].

\bibitem{Mustafa:2004dr}
M.~G. Mustafa, {\it {Energy loss of charm quarks in the quark-gluon plasma:
  Collisional versus radiative}},  {\em Phys. Rev. C} {\bf 72} (2005) 014905,
  [\href{http://arxiv.org/abs/hep-ph/0412402}{{\tt hep-ph/0412402}}].

\bibitem{Djordjevic:2006tw}
M.~Djordjevic, {\it {Collisional energy loss in a finite size QCD matter}},
  {\em Phys. Rev. C} {\bf 74} (2006) 064907,
  [\href{http://arxiv.org/abs/nucl-th/0603066}{{\tt nucl-th/0603066}}].

\bibitem{Wang:2006qr}
X.-N. Wang, {\it {Interference effect in elastic parton energy loss in a finite
  medium}},  {\em Phys. Lett. B} {\bf 650} (2007) 213--218,
  [\href{http://arxiv.org/abs/nucl-th/0604040}{{\tt nucl-th/0604040}}].

\bibitem{Schenke:2008gg}
B.~Schenke, M.~Strickland, A.~Dumitru, Y.~Nara, and C.~Greiner, {\it
  {Transverse momentum diffusion and jet energy loss in non-Abelian plasmas}},
  {\em Phys. Rev. C} {\bf 79} (2009) 034903,
  [\href{http://arxiv.org/abs/0810.1314}{{\tt arXiv:0810.1314}}].

\bibitem{PhysRevLett.100.072301}
G.-Y. Qin, J.~Ruppert, C.~Gale, S.~Jeon, G.~D. Moore, and M.~G. Mustafa, {\it
  Radiative and collisional jet energy loss in the quark-gluon plasma at the
  bnl relativistic heavy ion collider},  {\em Phys. Rev. Lett.} {\bf 100} (Feb,
  2008) 072301.

\bibitem{Liu:2006ug}
H.~Liu, K.~Rajagopal, and U.~A. Wiedemann, {\it {Calculating the jet quenching
  parameter from AdS/CFT}},  {\em Phys. Rev. Lett.} {\bf 97} (2006) 182301,
  [\href{http://arxiv.org/abs/hep-ph/0605178}{{\tt hep-ph/0605178}}].

\bibitem{Wang:2002ri}
E.~Wang and X.-N. Wang, {\it {Jet tomography of dense and nuclear matter}},
  {\em Phys. Rev. Lett.} {\bf 89} (2002) 162301,
  [\href{http://arxiv.org/abs/hep-ph/0202105}{{\tt hep-ph/0202105}}].

\bibitem{Idilbi:2008vm}
A.~Idilbi and A.~Majumder, {\it {Extending Soft-Collinear-Effective-Theory to
  describe hard jets in dense QCD media}},  {\em Phys. Rev. D} {\bf 80} (2009)
  054022, [\href{http://arxiv.org/abs/0808.1087}{{\tt arXiv:0808.1087}}].

\bibitem{Ovanesyan:2011xy}
G.~Ovanesyan and I.~Vitev, {\it {An effective theory for jet propagation in
  dense QCD matter: jet broadening and medium-induced bremsstrahlung}},  {\em
  JHEP} {\bf 06} (2011) 080, [\href{http://arxiv.org/abs/1103.1074}{{\tt
  arXiv:1103.1074}}].

\bibitem{Chien:2015vja}
Y.-T. Chien, A.~Emerman, Z.-B. Kang, G.~Ovanesyan, and I.~Vitev, {\it {Jet
  Quenching from QCD Evolution}},  {\em Phys. Rev. D} {\bf 93} (2016), no.~7
  074030, [\href{http://arxiv.org/abs/1509.02936}{{\tt arXiv:1509.02936}}].

\bibitem{Cao:2017qpx}
S.~Cao and A.~Majumder, {\it {Nuclear modification of leading hadrons and jets
  within a virtuality ordered parton shower}},  {\em Phys. Rev. C} {\bf 101}
  (2020), no.~2 024903, [\href{http://arxiv.org/abs/1712.10055}{{\tt
  arXiv:1712.10055}}].

\bibitem{Arnold:2002zm}
P.~B. Arnold, G.~D. Moore, and L.~G. Yaffe, {\it {Effective kinetic theory for
  high temperature gauge theories}},  {\em JHEP} {\bf 01} (2003) 030,
  [\href{http://arxiv.org/abs/hep-ph/0209353}{{\tt hep-ph/0209353}}].

\bibitem{Arnold:2003zc}
P.~B. Arnold, G.~D. Moore, and L.~G. Yaffe, {\it {Transport coefficients in
  high temperature gauge theories. 2. Beyond leading log}},  {\em JHEP} {\bf
  05} (2003) 051, [\href{http://arxiv.org/abs/hep-ph/0302165}{{\tt
  hep-ph/0302165}}].

\bibitem{Jeon:2003gi}
S.~Jeon and G.~D. Moore, {\it {Energy loss of leading partons in a thermal QCD
  medium}},  {\em Phys. Rev. C} {\bf 71} (2005) 034901,
  [\href{http://arxiv.org/abs/hep-ph/0309332}{{\tt hep-ph/0309332}}].

\bibitem{Schenke:2009gb}
B.~Schenke, C.~Gale, and S.~Jeon, {\it {MARTINI: An Event generator for
  relativistic heavy-ion collisions}},  {\em Phys. Rev. C} {\bf 80} (2009)
  054913, [\href{http://arxiv.org/abs/0909.2037}{{\tt arXiv:0909.2037}}].

\bibitem{Ghiglieri:2015ala}
J.~Ghiglieri, G.~D. Moore, and D.~Teaney, {\it {Jet-Medium Interactions at NLO
  in a Weakly-Coupled Quark-Gluon Plasma}},  {\em JHEP} {\bf 03} (2016) 095,
  [\href{http://arxiv.org/abs/1509.07773}{{\tt arXiv:1509.07773}}].

\bibitem{He:2015pra}
Y.~He, T.~Luo, X.-N. Wang, and Y.~Zhu, {\it {Linear Boltzmann Transport for Jet
  Propagation in the Quark-Gluon Plasma: Elastic Processes and Medium Recoil}},
   {\em Phys. Rev. C} {\bf 91} (2015) 054908,
  [\href{http://arxiv.org/abs/1503.03313}{{\tt arXiv:1503.03313}}]. [Erratum:
  Phys.Rev.C 97, 019902 (2018)].

\bibitem{Cao:2017hhk}
S.~Cao, T.~Luo, G.-Y. Qin, and X.-N. Wang, {\it {Heavy and light flavor jet
  quenching at RHIC and LHC energies}},  {\em Phys. Lett. B} {\bf 777} (2018)
  255--259, [\href{http://arxiv.org/abs/1703.00822}{{\tt arXiv:1703.00822}}].

\bibitem{He:2018xjv}
Y.~He, S.~Cao, W.~Chen, T.~Luo, L.-G. Pang, and X.-N. Wang, {\it {Interplaying
  mechanisms behind single inclusive jet suppression in heavy-ion collisions}},
   {\em Phys. Rev. C} {\bf 99} (2019), no.~5 054911,
  [\href{http://arxiv.org/abs/1809.02525}{{\tt arXiv:1809.02525}}].

\bibitem{Ke:2018jem}
W.~Ke, Y.~Xu, and S.~A. Bass, {\it {Modified Boltzmann approach for modeling
  the splitting vertices induced by the hot QCD medium in the deep
  Landau-Pomeranchuk-Migdal region}},  {\em Phys. Rev. C} {\bf 100} (2019),
  no.~6 064911, [\href{http://arxiv.org/abs/1810.08177}{{\tt
  arXiv:1810.08177}}].

\bibitem{Chen:2017zte}
W.~Chen, S.~Cao, T.~Luo, L.-G. Pang, and X.-N. Wang, {\it {Effects of
  jet-induced medium excitation in $\gamma$-hadron correlation in A+A
  collisions}},  {\em Phys. Lett. B} {\bf 777} (2018) 86--90,
  [\href{http://arxiv.org/abs/1704.03648}{{\tt arXiv:1704.03648}}].

\bibitem{PhysRevLett.120.232001}
P.~Caucal, E.~Iancu, A.~H. Mueller, and G.~Soyez, {\it Vacuumlike jet
  fragmentation in a dense qcd medium},  {\em Phys. Rev. Lett.} {\bf 120} (Jun,
  2018) 232001.

\bibitem{Majumder:2013re}
A.~Majumder, {\it {Incorporating Space-Time Within Medium-Modified Jet Event
  Generators}},  {\em Phys. Rev. C} {\bf 88} (2013) 014909,
  [\href{http://arxiv.org/abs/1301.5323}{{\tt arXiv:1301.5323}}].

\bibitem{Wang:2001ifa}
X.-N. Wang and X.-f. Guo, {\it {Multiple parton scattering in nuclei: Parton
  energy loss}},  {\em Nucl. Phys. A} {\bf 696} (2001) 788--832,
  [\href{http://arxiv.org/abs/hep-ph/0102230}{{\tt hep-ph/0102230}}].

\bibitem{Majumder:2009ge}
A.~Majumder, {\it {Hard collinear gluon radiation and multiple scattering in a
  medium}},  {\em Phys. Rev. D} {\bf 85} (2012) 014023,
  [\href{http://arxiv.org/abs/0912.2987}{{\tt arXiv:0912.2987}}].

\bibitem{Wang:2009qb}
W.-t. Deng and X.-N. Wang, {\it {Multiple Parton Scattering in Nuclei: Modified
  DGLAP Evolution for Fragmentation Functions}},  {\em Phys. Rev. C} {\bf 81}
  (2010) 024902, [\href{http://arxiv.org/abs/0910.3403}{{\tt
  arXiv:0910.3403}}].

\bibitem{Cao:2017zih}
{\bf JETSCAPE} Collaboration, S.~Cao et~al., {\it {Multistage Monte-Carlo
  simulation of jet modification in a static medium}},  {\em Phys. Rev. C} {\bf
  96} (2017), no.~2 024909, [\href{http://arxiv.org/abs/1705.00050}{{\tt
  arXiv:1705.00050}}].

\bibitem{Kauder:2018cdt}
{\bf JETSCAPE} Collaboration, K.~Kauder, {\it {JETSCAPE v1.0 Quickstart
  Guide}},  {\em Nucl. Phys. A} {\bf 982} (2019) 615--618,
  [\href{http://arxiv.org/abs/1807.09615}{{\tt arXiv:1807.09615}}].

\bibitem{Putschke:2019yrg}
J.~Putschke et~al., {\it {The JETSCAPE framework}},
  \href{http://arxiv.org/abs/1903.07706}{{\tt arXiv:1903.07706}}.

\bibitem{Soltz:2019aea}
{\bf Jetscape} Collaboration, R.~Soltz, {\it {Bayesian extraction of $\hat{q}$
  with multi-stage jet evolution approach}},  {\em PoS} {\bf HardProbes2018}
  (2019) 048.

\bibitem{Park:2020mkl}
{\bf JETSCAPE} Collaboration, C.~Park et~al., {\it {Constraints on jet
  quenching from a multi-stage energy-loss approach}},  in {\em {10th
  International Conference on Hard and Electromagnetic Probes of High-Energy
  Nuclear Collisions}: {Hard Probes 2020~}}, 9, 2020.
\newblock \href{http://arxiv.org/abs/2009.02410}{{\tt arXiv:2009.02410}}.

\bibitem{Dorau:2019ozd}
P.~Dorau, J.-B. Rose, D.~Pablos, and H.~Elfner, {\it {Jet Quenching in the
  Hadron Gas: An Exploratory Study}},  {\em Phys. Rev. C} {\bf 101} (2020),
  no.~3 035208, [\href{http://arxiv.org/abs/1910.07027}{{\tt
  arXiv:1910.07027}}].

\bibitem{Tachibana:2017syd}
Y.~Tachibana, N.-B. Chang, and G.-Y. Qin, {\it {Full jet in quark-gluon plasma
  with hydrodynamic medium response}},  {\em Phys. Rev. C} {\bf 95} (2017),
  no.~4 044909, [\href{http://arxiv.org/abs/1701.07951}{{\tt
  arXiv:1701.07951}}].

\bibitem{Casalderrey-Solana:2016jvj}
J.~Casalderrey-Solana, D.~Gulhan, G.~Milhano, D.~Pablos, and K.~Rajagopal, {\it
  {Angular Structure of Jet Quenching Within a Hybrid Strong/Weak Coupling
  Model}},  {\em JHEP} {\bf 03} (2017) 135,
  [\href{http://arxiv.org/abs/1609.05842}{{\tt arXiv:1609.05842}}].

\bibitem{Casalderrey-Solana:2020rsj}
J.~Casalderrey-Solana, J.~G. Milhano, D.~Pablos, K.~Rajagopal, and X.~Yao, {\it
  {Jet Wake from Linearized Hydrodynamics}},
  \href{http://arxiv.org/abs/2010.01140}{{\tt arXiv:2010.01140}}.

\bibitem{KunnawalkamElayavalli:2017hxo}
R.~Kunnawalkam~Elayavalli and K.~C. Zapp, {\it {Medium response in JEWEL and
  its impact on jet shape observables in heavy ion collisions}},  {\em JHEP}
  {\bf 07} (2017) 141, [\href{http://arxiv.org/abs/1707.01539}{{\tt
  arXiv:1707.01539}}].

\bibitem{Sjostrand:2014zea}
T.~Sjöstrand, S.~Ask, J.~R. Christiansen, R.~Corke, N.~Desai, P.~Ilten,
  S.~Mrenna, S.~Prestel, C.~O. Rasmussen, and P.~Z. Skands, {\it {An
  introduction to PYTHIA 8.2}},  {\em Comput. Phys. Commun.} {\bf 191} (2015)
  159--177, [\href{http://arxiv.org/abs/1410.3012}{{\tt arXiv:1410.3012}}].

\bibitem{ATL-PHYS-PUB-2014-021}
{\bf ATLAS} Collaboration, {\it {ATLAS Pythia 8 tunes to 7 TeV datas}},  Tech.
  Rep. ATL-PHYS-PUB-2014-021, CERN, Geneva, Nov, 2014.

\bibitem{Pumplin:2002vw}
J.~Pumplin, D.~Stump, J.~Huston, H.~Lai, P.~M. Nadolsky, and W.~Tung, {\it {New
  generation of parton distributions with uncertainties from global QCD
  analysis}},  {\em JHEP} {\bf 07} (2002) 012,
  [\href{http://arxiv.org/abs/hep-ph/0201195}{{\tt hep-ph/0201195}}].

\bibitem{Eskola:2009uj}
K.~Eskola, H.~Paukkunen, and C.~Salgado, {\it {EPS09: A New Generation of NLO
  and LO Nuclear Parton Distribution Functions}},  {\em JHEP} {\bf 04} (2009)
  065, [\href{http://arxiv.org/abs/0902.4154}{{\tt arXiv:0902.4154}}].

\bibitem{PhysRevD.27.140}
J.~D. Bjorken, {\it Highly relativistic nucleus-nucleus collisions: The central
  rapidity region},  {\em Phys. Rev. D} {\bf 27} (Jan, 1983) 140--151.

\bibitem{Arnold:2008zu}
P.~B. Arnold and C.~Dogan, {\it {QCD Splitting/Joining Functions at Finite
  Temperature in the Deep LPM Regime}},  {\em Phys. Rev. D} {\bf 78} (2008)
  065008, [\href{http://arxiv.org/abs/0804.3359}{{\tt arXiv:0804.3359}}].

\bibitem{CaronHuot:2010bp}
S.~Caron-Huot and C.~Gale, {\it {Finite-size effects on the radiative energy
  loss of a fast parton in hot and dense strongly interacting matter}},  {\em
  Phys. Rev. C} {\bf 82} (2010) 064902,
  [\href{http://arxiv.org/abs/1006.2379}{{\tt arXiv:1006.2379}}].

\bibitem{Baier:1998yf}
R.~Baier, Y.~L. Dokshitzer, A.~H. Mueller, and D.~Schiff, {\it {Radiative
  energy loss of high-energy partons traversing an expanding QCD plasma}},
  {\em Phys. Rev. C} {\bf 58} (1998) 1706--1713,
  [\href{http://arxiv.org/abs/hep-ph/9803473}{{\tt hep-ph/9803473}}].

\bibitem{PhysRevD.10.186}
F.~Cooper and G.~Frye, {\it Single-particle distribution in the hydrodynamic
  and statistical thermodynamic models of multiparticle production},  {\em
  Phys. Rev. D} {\bf 10} (Jul, 1974) 186--189.

\bibitem{Bazavov:2014pvz}
{\bf HotQCD} Collaboration, A.~Bazavov et~al., {\it {Equation of state in
  (2+1)-flavor QCD}},  {\em Phys. Rev.} {\bf D90} (2014) 094503,
  [\href{http://arxiv.org/abs/1407.6387}{{\tt arXiv:1407.6387}}].

\bibitem{Cacciari:2008gp}
M.~Cacciari, G.~P. Salam, and G.~Soyez, {\it {The anti-$k_t$ jet clustering
  algorithm}},  {\em JHEP} {\bf 04} (2008) 063,
  [\href{http://arxiv.org/abs/0802.1189}{{\tt arXiv:0802.1189}}].

\bibitem{Cacciari:2011ma}
M.~Cacciari, G.~P. Salam, and G.~Soyez, {\it {FastJet User Manual}},  {\em Eur.
  Phys. J. C} {\bf 72} (2012) 1896, [\href{http://arxiv.org/abs/1111.6097}{{\tt
  arXiv:1111.6097}}].

\bibitem{Bernhard:2016tnd}
J.~E. Bernhard, J.~S. Moreland, S.~A. Bass, J.~Liu, and U.~Heinz, {\it
  {Applying Bayesian parameter estimation to relativistic heavy-ion collisions:
  simultaneous characterization of the initial state and quark-gluon plasma
  medium}},  {\em Phys. Rev. C} {\bf 94} (2016), no.~2 024907,
  [\href{http://arxiv.org/abs/1605.03954}{{\tt arXiv:1605.03954}}].

\bibitem{Aaboud:2018twu}
{\bf ATLAS} Collaboration, M.~Aaboud et~al., {\it {Measurement of the nuclear
  modification factor for inclusive jets in Pb+Pb collisions at
  $\sqrt{s_\mathrm{NN}}=5.02$ TeV with the ATLAS detector}},  {\em Phys. Lett.
  B} {\bf 790} (2019) 108--128, [\href{http://arxiv.org/abs/1805.05635}{{\tt
  arXiv:1805.05635}}].

\bibitem{Acharya:2019jyg}
{\bf ALICE} Collaboration, S.~Acharya et~al., {\it {Measurements of inclusive
  jet spectra in pp and central Pb-Pb collisions at $\sqrt{s_{\rm{NN}}}$ = 5.02
  TeV}},  {\em Phys. Rev. C} {\bf 101} (2020), no.~3 034911,
  [\href{http://arxiv.org/abs/1909.09718}{{\tt arXiv:1909.09718}}].

\bibitem{Adam:2020wen}
{\bf STAR} Collaboration, J.~Adam et~al., {\it {Measurement of inclusive
  charged-particle jet production in Au+Au collisions at $\sqrt{s_{NN}}$=200
  GeV}},  \href{http://arxiv.org/abs/2006.00582}{{\tt arXiv:2006.00582}}.

\bibitem{CMS-PAS-HIN-18-014}
{\bf CMS Collaboration} Collaboration, {\it {Measurement of Jet Nuclear
  Modification Factor in PbPb Collisions at $\sqrt{s_{NN}}$ = 5.02 TeV with
  CMS}},  Tech. Rep. CMS-PAS-HIN-18-014, CERN, Geneva, 2019.

\bibitem{Khachatryan:2016odn}
{\bf CMS} Collaboration, V.~Khachatryan et~al., {\it {Charged-particle nuclear
  modification factors in PbPb and pPb collisions at $
  \sqrt{s_{\mathrm{N}\;\mathrm{N}}}=5.02 $ TeV}},  {\em JHEP} {\bf 04} (2017)
  039, [\href{http://arxiv.org/abs/1611.01664}{{\tt arXiv:1611.01664}}].

\bibitem{Sirunyan:2017xss}
{\bf CMS} Collaboration, A.~M. Sirunyan et~al., {\it {Nuclear modification
  factor of D$^0$ mesons in PbPb collisions at $\sqrt{s_\mathrm{NN}} = 5.02$
  TeV}},  {\em Phys. Lett. B} {\bf 782} (2018) 474--496,
  [\href{http://arxiv.org/abs/1708.04962}{{\tt arXiv:1708.04962}}].

\bibitem{Adare:2012wg}
{\bf PHENIX} Collaboration, A.~Adare et~al., {\it {Neutral pion production with
  respect to centrality and reaction plane in Au$+$Au collisions at
  $\sqrt{s_{NN}}$=200 GeV}},  {\em Phys. Rev. C} {\bf 87} (2013), no.~3 034911,
  [\href{http://arxiv.org/abs/1208.2254}{{\tt arXiv:1208.2254}}].

\bibitem{Fries:2003vb}
R.~Fries, B.~Muller, C.~Nonaka, and S.~Bass, {\it {Hadronization in heavy ion
  collisions: Recombination and fragmentation of partons}},  {\em Phys. Rev.
  Lett.} {\bf 90} (2003) 202303,
  [\href{http://arxiv.org/abs/nucl-th/0301087}{{\tt nucl-th/0301087}}].

\bibitem{Acharya:2018hre}
{\bf ALICE} Collaboration, S.~Acharya et~al., {\it {Measurement of D$^{0}$,
  D$^{+}$, D$^{*+}$ and D$_{s}^{+}$ production in Pb-Pb collisions at $
  \sqrt{{\mathrm{s}}_{\mathrm{NN}}}=5.02 $ TeV}},  {\em JHEP} {\bf 10} (2018)
  174, [\href{http://arxiv.org/abs/1804.09083}{{\tt arXiv:1804.09083}}].

\bibitem{Blaizot:2013hx}
J.-P. Blaizot, E.~Iancu, and Y.~Mehtar-Tani, {\it {Medium-induced QCD cascade:
  democratic branching and wave turbulence}},  {\em Phys. Rev. Lett.} {\bf 111}
  (2013) 052001, [\href{http://arxiv.org/abs/1301.6102}{{\tt
  arXiv:1301.6102}}].

\bibitem{Blaizot:2014ula}
J.-P. Blaizot, Y.~Mehtar-Tani, and M.~A.~C. Torres, {\it {Angular structure of
  the in-medium QCD cascade}},  {\em Phys. Rev. Lett.} {\bf 114} (2015), no.~22
  222002, [\href{http://arxiv.org/abs/1407.0326}{{\tt arXiv:1407.0326}}].

\bibitem{Fister:2014zxa}
L.~Fister and E.~Iancu, {\it {Medium-induced jet evolution: wave turbulence and
  energy loss}},  {\em JHEP} {\bf 03} (2015) 082,
  [\href{http://arxiv.org/abs/1409.2010}{{\tt arXiv:1409.2010}}].

\bibitem{Blaizot:2014rla}
J.-P. Blaizot, L.~Fister, and Y.~Mehtar-Tani, {\it {Angular distribution of
  medium-induced QCD cascades}},  {\em Nucl. Phys. A} {\bf 940} (2015) 67--88,
  [\href{http://arxiv.org/abs/1409.6202}{{\tt arXiv:1409.6202}}].

\bibitem{Blaizot:2015jea}
J.-P. Blaizot and Y.~Mehtar-Tani, {\it {Energy flow along the medium-induced
  parton cascade}},  {\em Annals Phys.} {\bf 368} (2016) 148--176,
  [\href{http://arxiv.org/abs/1501.03443}{{\tt arXiv:1501.03443}}].

\bibitem{Schlichting:2020lef}
S.~Schlichting and I.~Soudi, {\it {Fragmentation and equilibration of jets in a
  QCD plasma}},  \href{http://arxiv.org/abs/2008.04928}{{\tt
  arXiv:2008.04928}}.

\bibitem{Ke:2020coj}
W.~Ke, {\em {Partonic transport model application to heavy flavor}}.
\newblock Phd thesis, Duke University, 2019.
\newblock \href{http://arxiv.org/abs/2001.02766}{{\tt arXiv:2001.02766}}.

\bibitem{Spousta:2015fca}
M.~Spousta and B.~Cole, {\it {Interpreting single jet measurements in Pb $+$ Pb
  collisions at the LHC}},  {\em Eur. Phys. J. C} {\bf 76} (2016), no.~2 50,
  [\href{http://arxiv.org/abs/1504.05169}{{\tt arXiv:1504.05169}}].

\bibitem{Aaboud:2018hpb}
{\bf ATLAS} Collaboration, M.~Aaboud et~al., {\it {Measurement of jet
  fragmentation in Pb+Pb and $pp$ collisions at $\sqrt{s_{NN}} = 5.02$ TeV with
  the ATLAS detector}},  {\em Phys. Rev. C} {\bf 98} (2018), no.~2 024908,
  [\href{http://arxiv.org/abs/1805.05424}{{\tt arXiv:1805.05424}}].

\bibitem{Sirunyan:2018jqr}
{\bf CMS} Collaboration, A.~M. Sirunyan et~al., {\it {Jet properties in PbPb
  and pp collisions at $ \sqrt{s_{\mathrm{N}\;\mathrm{N}}}=5.02 $ TeV}},  {\em
  JHEP} {\bf 05} (2018) 006, [\href{http://arxiv.org/abs/1803.00042}{{\tt
  arXiv:1803.00042}}].

\bibitem{Bernhard:2018hnz}
J.~E. Bernhard, {\em {Bayesian parameter estimation for relativistic heavy-ion
  collisions}}.
\newblock PhD thesis, Duke U., 4, 2018.
\newblock \href{http://arxiv.org/abs/1804.06469}{{\tt arXiv:1804.06469}}.

\bibitem{Moreland:2014oya}
J.~S. Moreland, J.~E. Bernhard, and S.~A. Bass, {\it {Alternative ansatz to
  wounded nucleon and binary collision scaling in high-energy nuclear
  collisions}},  {\em Phys. Rev. C} {\bf 92} (2015), no.~1 011901,
  [\href{http://arxiv.org/abs/1412.4708}{{\tt arXiv:1412.4708}}].

\bibitem{Broniowski:2008qk}
W.~Broniowski, W.~Florkowski, M.~Chojnacki, and A.~Kisiel, {\it {Free-streaming
  approximation in early dynamics of relativistic heavy-ion collisions}},  {\em
  Phys. Rev. C} {\bf 80} (2009) 034902,
  [\href{http://arxiv.org/abs/0812.3393}{{\tt arXiv:0812.3393}}].

\bibitem{Song:2007ux}
H.~Song and U.~W. Heinz, {\it {Causal viscous hydrodynamics in 2+1 dimensions
  for relativistic heavy-ion collisions}},  {\em Phys. Rev. C} {\bf 77} (2008)
  064901, [\href{http://arxiv.org/abs/0712.3715}{{\tt arXiv:0712.3715}}].

\bibitem{SHEN201661}
C.~Shen, Z.~Qiu, H.~Song, J.~Bernhard, S.~Bass, and U.~Heinz, {\it The
  iebe-vishnu code package for relativistic heavy-ion collisions},  {\em
  Computer Physics Communications} {\bf 199} (2016) 61 -- 85.

\bibitem{Bjorken:1982qr}
J.~Bjorken, {\it {Highly Relativistic Nucleus-Nucleus Collisions: The Central
  Rapidity Region}},  {\em Phys. Rev. D} {\bf 27} (1983) 140--151.

\bibitem{R}
{R Core Team}, {\em R: A Language and Environment for Statistical Computing}.
\newblock R Foundation for Statistical Computing, Vienna, Austria, 2019.

\bibitem{R-LHS}
R.~Carnell, {\em lhs: Latin Hypercube Samples}, 2019.
\newblock R package version 1.0.1.

\bibitem{scikit-learn}
F.~Pedregosa, G.~Varoquaux, A.~Gramfort, V.~Michel, B.~Thirion, O.~Grisel,
  M.~Blondel, P.~Prettenhofer, R.~Weiss, V.~Dubourg, J.~Vanderplas, A.~Passos,
  D.~Cournapeau, M.~Brucher, M.~Perrot, and E.~Duchesnay, {\it Scikit-learn:
  Machine learning in {P}ython},  {\em Journal of Machine Learning Research}
  {\bf 12} (2011) 2825--2830.

\bibitem{emcee}
D.~{Foreman-Mackey}, D.~W. {Hogg}, D.~{Lang}, and J.~{Goodman}, {\it {emcee:
  The MCMC Hammer}},  {\em {Publications of the Astronomical Society of the
  Pacific}} {\bf 125} (Mar., 2013) 306,
  [\href{http://arxiv.org/abs/1202.3665}{{\tt arXiv:1202.3665}}].

\end{thebibliography}\endgroup
\end{document}